\documentstyle[fleqn,12pt]{article}
\begin{document}
\renewcommand{\theequation}{\arabic{section}.\arabic{equation}}
\renewcommand{\thesection}{\S\arabic{section}}
\baselineskip=20pt
%\begin{titlepage}
\noindent \hspace*{7.5cm}NIM-TH 95-11\\

\noindent\hspace*{7.5cm}\parbox{7cm}{Talk at the second pacific winter School
in
Theoretical Physics. Korea, 1995}\\
\vspace*{2cm}

\begin{center}
{\huge   \bf        Introduction to {\em Yangian} in Physics}
\end{center}
\vspace*{0.7cm}

\begin{center}
{\bf                   Mo-Lin Ge\dag, Kang Xue\ddag\ and Yiwen Wang\dag}\\
\vspace*{0.5cm}
{\small {\it          \dag Theoretical Physics Division,
                       Nankai Institute of Mathematics,}}\\
{\small {\it           Tianjin 300071, People's Republic of China}}\\
{\small {\it     \ddag  Physics Department, Northeast Normal University,}}\\
{\small {\it           Changchun 130024, People's Republic of China }}
\end{center}

\vspace*{0.3in}

\begin{abstract}
\baselineskip=20pt
This is an introduction to the physical pictures of {\em Yangian} symmetry.
All the discussions are based on the RTT relations which have been known to
be related to the Hamiltonian formulations for quantum integrable systems.
The explicit calculations of {\em Yangian} associated with $gl(2)$ are given
through solving the RTT relations. It contains the Drinfeld's $Y(sl(2))$,
$Y(sl(n))$ and Heisenberg type of Hopf algebras. Examples are emphasized to
show the physical pictures, such as Hubbard model, Haldane-Shastry model,
Long-range interaction models and Goryachev-Chaplygin(G-C) gyrostat. The
example of the trigonometric extension of G-C gyrostat are also shown
that leads to the non-commutative geometry naturally on the basis
of the truncated $q$-affine algebra.
%\noindent PACS numbers: 05.30.-d, 03.65.Fd, 05.50.+q
\end{abstract}
%\end{titlepage}
\newpage
\baselineskip=16pt
\section{Introduction}
\setcounter{equation}{0}

An integrable system is said to be solved if the equation
of motion is solved or all the conserved quantities are found. From the
point of view of N\"other theorem equations of motion and conserved
quantities co-exist in connecting with certain symmetries of the
${\cal L}$agrangian.
In quantum mechanics, to obtain the spectrum we can either solve the
Sch\"odinger equation or find all the symmetries in the system. The typical
example is the Hydrogen atom which possesses another SU(2) invariance due to
the Runge-Lenz vector besides the well-known SU(2) associated with the
spherically symmetric potential U(r). With the given SO(4)=SU(2)$\otimes$SU(2))
symmetry the spectrum of a Hydrogen atom can easily be obtained without the
explicit form of the wave function satisfying Schr\"odinger
equation~\cite{re1}.
In this example the family of the symmetries consist in the two
SU(2) that commute with each other, namely, the two conserved quantum
operators corresponding to the symmetries form a commuting family. This
simple example provides a representation of completely quantum integrable
systems.

To extend the above analysis to nonlinear models it should be noticed that
a quantum many-body problem in the first quantization form with conserved
particle number $N$ is, in principle, equivalent to the corresponding
quantum field theory(second quantization form). The typical evidence
is the equivalence between $\delta$-interaction model with the Hamiltonian
\begin{equation}
H=-\sum^N_{i=1}\partial ^2_i+2c \sum_{i\neq j}\delta(x_i-x_j)\label{p1.1}
\end{equation}
and the nonlinear Schr\"odinger equation~\cite{re4}:
\begin{equation}
{\hat H}=\int dx\left(\frac 12 \partial_x\phi \partial_x\phi^*+2c(\phi
\phi^*)^2
\right) \label{p1.2}
\end{equation}
where $\phi=\phi(x,t)$ stands for the second quantized operator.

In solving a many-body problem we meet also two approaches. The first
one is to solve the Schr\"odinger equation directly. It is the well-known
Bethe-Anastz methods~\cite{re2}. The second approach is to find the commuting
family of conserved quantities consisting of quantum operators, that is
known as the Quantum Inverse Scattering Methods(QISM)~\cite{re3}. Although
the Bethe-Ansatz approach is more powerful in solving the 1-dim.
many-body problems, the QISM provides a model-independently systematic method
to deal with the (1+1) dimensional integrable QFT~\cite{re3}.
In this approach the RTT relation plays the central role. For a wide class of
models it takes the form~\cite{re5}:
\begin{equation}
R(u-v)(T(u)\otimes I)(I\otimes T(v))=(I\otimes T(v))(T(u)\otimes I)R(u-v)
\label{p1.3}
\end{equation}
where $T(u)$ stands for the quantum transfer matrix and $u$ and $v$ the
spectral parameters. In general,
\begin{equation}
T(u)=||T_{ab}(u)||^N_{a,b=1}\label{p1.4}
\end{equation}
and $R$ is then $N^2$ by $N^2$ c-numbered matrix. In eq.(\ref{p1.4}) each
element $T_{ab}(u)$ is $u$-dependent quantum operator. For this reason the
space spanned by the indices $a,b,\ldots$ is called ``auxiliary space'',
whereas the Hilbert space spanned by $T_{ab}$ itself is called ``quantum
space''.

Defining
\begin{equation}
\stackrel{(1)}{T}(u)=T(u)\otimes I,\ \  \stackrel{(2)}{T}(v)=I\otimes T(v)
\label{p1.5}
\end{equation}
the  eq.(\ref{p1.5}) can be recast into the form~\cite{re5}
\begin{equation}
R(u-v)\stackrel{(1)}{T}(u) \stackrel{(2)}{T}(v)=
\stackrel{(2)}{T}(v) \stackrel{(1)}{T}(u)R(u-v)\label{p1.6}
\end{equation}
where each element of $T$ is a quantum operator so that the ordering for
$\stackrel{(1)}{T}$ and $\stackrel{(2)}{T}$ is very important. Noting that
\[
(\stackrel{(1)}{T}(u) \stackrel{(2)}{T}(v))_{ij,kl}
=(T(u)\otimes I)_{ij,mn} (I\otimes T(v)_{mn,kl}
=(T(u)\otimes T(v))_{ij,kl}
\]
the component form of eq.~({\ref{p1.6}) is
\[
R(u-v)_{ij,ms}(T(u)\otimes T(v))_{ms,kl}=(T(v)\otimes
T(u))_{ji,sr}R(u-v)_{rs,kl}
\]
that can be rewritten in the form
\[
\check{R}(u-v)_{ij,rs}(T(u)\otimes T(v))_{rs,kl}=(T(v)\otimes T(u))_{ij,rs}
\check{R}(u-v)_{rs,kl}
\]
i.e.,
\begin{equation}
\check{R}(u-v)(T(u)\otimes T(v))=(T(v)\otimes T(u))\check{R}(u-v) \label{p1.7}
\end{equation}
where
\begin{equation}
\check{R}(u)=PR(u)\ \ \mbox{or}\ \ \check{R}_{ij,kl}=R_{ji,kl} \label{p1.8}
\end{equation}
{}From eq.~({\ref{p1.6}) or eq.~(\ref{p1.7}) it follows:\\

\noindent {\bf (1)} Because $\check{R}(u)$ is non-singular matrix:
\begin{equation}
T(u)\otimes T(v)={\check{R}}^{-1}(u-v)(T(v)\otimes
T(u))\check{R}(u-v)\label{p1.9}
\end{equation}
hence, by taking the trace of both of sides we have
\begin{equation}
[\mbox{tr}T(u),\ \mbox{tr}T(v)]=0 \label{p1.10}
\end{equation}
where the commutation relation is taken over the quantum space. By taking the
expansion in terms of $u$ and $v$
\begin{equation}
\mbox{tr}T(u)=\sum u^{-n}J^{(n)}\label{p1.11}
\end{equation}
we get
\begin{equation}
[J^{(n)}, J^{(m)}]=0 \label{p1.12}
\end{equation}
i.e., there exists an infinite set of conserved quantum operators $J^{(n)}$.\\

\noindent{\bf (2)} In eq.~(\ref{p1.6}) the R-matrix gives rise to the exchange
between the space 1 and space 2, we can denote the $R$ by $R_{12}$. Similarly
it can be introduce $R_{ij}$ through
\[ R_{ij}\stackrel{(i)}{T} \stackrel{(j)}{T}=
\stackrel{(j)}{T} \stackrel{(i)}{T}R_{ij} \]
then
\[  R_{12}R_{13}R_{23}\stackrel{1}{T}\stackrel{2}{T}\stackrel{3}{T}=
\stackrel{3}{T}\stackrel{2}{T}\stackrel{1}{T}R_{12}R_{13}R_{23}\label{p1.13}
\]
and
\[
R_{23}R_{13}R_{12}\stackrel{1}{T}\stackrel{2}{T}\stackrel{3}{T}=
\stackrel{3}{T}\stackrel{2}{T}\stackrel{1}{T}R_{23}R_{13}R_{12}\label{p1.14}
\]
that lead to
\[
[R_{12}R_{13}R_{23}(R_{23}R_{13}R_{23})^{-1},\stackrel{3}{T}
\stackrel{2}{T}\stackrel{1}{T} ]=0 \label{p1.15}
\]
where the first term in the commutator should be unity because its
determinant is unity. Thus base on the associativity we have the
Yang-Baxter equation(YBE):
\begin{equation}
R_{12}(u_{12})R_{13}(u_{13})R_{23}(u_{23})=R_{23}(u_{23})R_{13}(u_{13})
R_{12}(u_{12})\ (u_{ij}=u_i-u_j) \label{p1.16}
\end{equation}
YBE is a set of algebraic equations. The component form of YBE reads:
\begin{equation}
R^{ij}_{rs}(u_{12}) R^{rk}_{lp}(u_{13}) R^{sp}_{lm}(u_{23})=
R^{jk}_{sp}(u_{23}) R^{ip}_{rn}(u_{13}) R^{rs}_{lm}(u_{12}) \label{p1.17}
\end{equation}
where the repeated indices mean summation.

Now the standard procedure to deal with a Y-B system is:
\begin{enumerate}
\item Solving the YBE to find a $R$-matrix satisfying eq.~(\ref{p1.17}).
\item Finding the commutation relations for $T_{ab}$ by solving the RTT
relation
eq.~(\ref{p1.6}) or eq.~(\ref{p1.7}).
\item Making a physical realization of $T_{ab}$.
\item Substituting the $T_{ab}$ into $\mbox{tr}T$ or other conserved quantities
that
commute with $\mbox{tr}T$ to find the Hamiltonian.
\item Following the QISM to find ``spectrum''.
\end{enumerate}
The step 2 is crucial. The RTT relation can give rise to new types of
symmetries associated with nonlinear interaction other than the Lie algebraic
structures for the linear models. The simple examples of $R$-matrices
satisfying YBE is given below:

\noindent{\bf (a)} The simplest rational form~\cite{re6}:
\begin{equation}
R(u)=u+P, \ \ \check{R}(u)=uP+I \label{p1.18}
\end{equation}
where $P$ is permutation operator $P(a\otimes b)=(b\otimes a)$. For a rational
$R$-matrix we have
\begin{equation}
T(u)=\sum^{\infty}_{n=0}u^{-n}T^{(n)} \label{p1.19}
\end{equation}
i.e., the expansion of the transfer matrix is taken in the ``half''
axis of the spectral parameter $u$.

\noindent{\bf (b)} The simplest trigonometric form~\cite{re7}:
\begin{eqnarray}
\check{R}(u)&=&\left[\begin{array}{cccc}xq-x^{-1}q^{-1}& & & \\
&q-q^{-1}&x-x^{-1}& \\ &x-x^{-1}&q-q^{-1}& \\ & & &xq-x^{-1}q^{-1}\end{array}
\right]\nonumber\\
&=&\left[\begin{array}{cccc} \sinh (u+\gamma) & & & \\ & \sinh \gamma &
\sinh u& \\ &\sinh u& \sinh \gamma & \\ & & &\sinh (u+\gamma)\end{array}
\right]\time 2 \label{p1.20}
\end{eqnarray}
where $x=e^u,\ q=e^{\gamma}$, or
\begin{eqnarray}
R(u)=\left[\begin{array}{cccc} \sinh (u+\gamma) & & & \\ & \sinh u &
\sinh \gamma& \\ &\sinh \gamma& \sinh u & \\ & & &\sinh (u+\gamma)\end{array}
\right] \label{p1.21}
\end{eqnarray}
that corresponds to the 6-vertex model. Obviously by $u\rightarrow \gamma u$
and let $\gamma\rightarrow 0$, eq.~(\ref{p1.20}) tends to
\[
\check{R}(u)\rightarrow \gamma(I+uP).
\]
Noting that the overall constant factor to $R$-matrix can always be dropped.
Therefore, eq.~(\ref{p1.18}) is nothing but the rational limit of
eq.~(\ref{p1.21}).

\noindent{\bf (c)} The simplest elliptic solution reads~\cite{re8}.
\begin{eqnarray}
R(u)=\left[\begin{array}{cccc} a(u) & & &d(u) \\ & b(u) & c(u)& \\
&c(u) & b(u) & \\ d(u)& & &a(u)\end{array}
\right] \label{p1.22}
\end{eqnarray}
where
\begin{eqnarray}
a(u) &=& \theta_0(\eta)\theta_0(u)\theta_1(u+\eta)\nonumber\\
b(u) &=& \theta_0(\eta)\theta_1(u)\theta_0(u+\eta)\nonumber\\
c(u) &=& \theta_1(\eta)\theta_0(u)\theta_0(u+\eta)\label{p1.23}\\
d(u) &=& \theta_1(\eta)\theta_1(u)\theta_1(u+\eta)\nonumber\\
\theta_0(u) &=& \prod^{\infty}_{n=1}(1-2p^{(n-1)/2}\cos 2\pi u+p^{2n-1})
(1-p^n) \nonumber\\
\theta_1(u) &=& 2p^{\frac 18}\sin \pi u\prod^{\infty}_{n=1}(1-2p^{n}
\cos 2\pi u+p^{2n})(1-p^n)\label{p1.24}
\end{eqnarray}
with $\eta$ and $p$ being free parameters, i.e., $R$-matrix given by
eq.~(\ref{p1.21}) is double periodic. When $p\approx 0$, $\theta_0(u)
\approx 2 \sin^2(\pi u),\ \theta_1(u)\approx 2 p^{\frac 18}\sin \pi u$,
we have $a(u)\approx f(p)\sin\pi(u+\eta),\ b(u)\approx f(p)\sin \pi u,\
c(u)\approx f(p) \sin \pi\eta$ and $d(u)\approx 0$. Hence, eq.(\ref{p1.22})
$\longrightarrow$ eq.(\ref{p1.21}), namely, it is reduced to the trigonometric
form.

The types (a), (b) and (c) correspond to the vertex models. Beyond them
there is the fourth type of $R$-matrix~\cite{re9}:

\noindent{\bf (d)} Chiral Potts  model

This type of solution is expressed in terms of functions $W_{pq}(a-b)$
and $\overline{W}_{pq}(a-b)$ that satisfy
\begin{eqnarray}
&&\sum_d W_{pr}(a-d)\overline{W}_{qr}(b-d)\overline{W}_{pq}(d-c)\nonumber\\
&=&R_{pqr}\overline{W}_{pr}(b-c)W_{qr}(a-c)W_{pq}(a-b) \label{p1.25}
\end{eqnarray}
where $W$ and $\overline{W}$ can be illustrated in terms of the diagrams

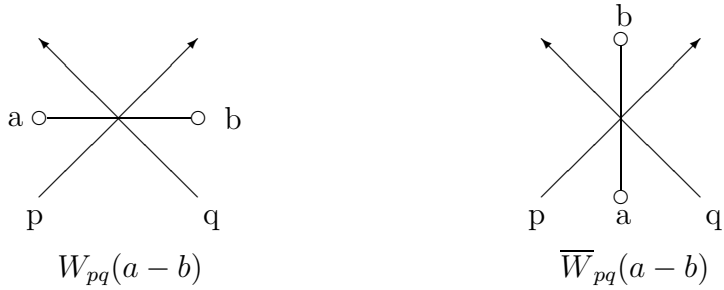
\begin{picture}(320,125)(-20,-15)
\put(30,20){\vector(1,1){60}}
\put(90,20){\vector(-1,1){60}}
\put(30,50){\circle{5}}
\put(90,50){\circle{5}}
\put(33,50){\line(1,0){54}}
\put(25,10){p}
\put(92,10){q}
\put(18,47){a}
\put(100,46){b}
\put(37,-10){$W_{pq}(a-b)$}
\put(220,20){\vector(1,1){60}}
\put(280,20){\vector(-1,1){60}}
\put(250,20){\circle{5}}
\put(250,80){\circle{5}}
\put(250,23){\line(0,1){54}}
\put(215,10){p}
\put(282,10){q}
\put(248,10){a}
\put(248,85){b}
\put(227,-10){$\overline{W}_{pq}(a-b)$}
\end{picture}\\
Eq.(\ref{p1.25}) can be solved:
\begin{eqnarray}
g_{pq}(n)&=&\frac{W_{pq}(n)}{W_{pq}(0)}=\prod^n_{j=1}\frac{d_pb_q-a_pc_q
\omega^j}{b_pd_q-c_pa_q\omega^j}\label{p1.26}\\
\overline{g}_{pq}(n)&=&\frac{\overline{W}_{pq}(n)}{\overline{W}_{pq}(0)}
=\prod^n_{j=1}\frac{\omega a_pd_q-d_pa_q\omega^j}{c_pb_q-b_pc_q\omega^j}
\label{p1.27}\\
\omega&=&e^{\frac{2\pi i}N}\nonumber
%\label{p1.28}
\end{eqnarray}
and
\begin{eqnarray}
R_{pqr}&=&\frac{f_{pq}f_{qr}}{f_{qr}}\label{p1.29}\\
f_{pq}&=&\left[\frac{\prod^N_{m=1}(\sum^N_{k=1}\omega^{mk}\overline{W}_{pq}
(k))}{\prod^N_{m=1}W_{pq}(m)}\right]^{\frac 1N}\label{p1.40}
\end{eqnarray}
The self-duality conditions:
\begin{equation}
c_p=d_p=1, \ \ a_p^N+b_p^N=0\label{p1.31}
\end{equation}
simplify the solution to
\begin{eqnarray}
g_{pq}(n)&=&\prod^n_{j=1}\frac{b_q-a_p\omega^j}{b_p-a_q\omega^j}\label{p1.32}\\
\overline{g}_{pq}(n)&=&\prod^n_{j=1}\frac{\omega a_p-a_q\omega^j}
{b_q-b_p\omega^j}\label{p1.33}
\end{eqnarray}
where $a_p,\ b_p$ and $\omega^j$ are all parameters. Especially, $a_p$ and
$b_p$ play the role as the spectral parameter $u$ in the vertex models.

The main goal of this lecture is to give a physical understanding on the
consequences of RTT relations for $R$-matrices are rational and trigonometric.
We shall first calculate the explicit commutation relations satisfied by
$T_{ab}\ (a,b=1,2)$ with the $R$-matrix given by the rational form
eq.~(\ref{p1.18}), then introduce the {\em Yangian} symmetry given by
Drinfeld~\cite{re10}. We try to give a phyiscal interpretation of Drinfeld's
theory. Next we would like to discuss variety of physical realizations
of {\em Yangian} and point out how the {\em Yangian} symmetry is used to
construct
Hamiltonian for a quantum integrable system. As an example, the long-range
interaction model for sl(n) will be discussed. The trigonometric solution
of $R$-matrix eq.~(\ref{p1.21}) gives rise to the quantum algebra. we shall
explain how to find a q-deformed model based on the truncated affine quantum
algebra. For reader's convenience we would like make the paper
self-contained.
\section{RTT Relation and Y(sl(2))}
\setcounter{equation}{0}

\subsection{RTT and its Commutation Relations}

For the given sl(2) $R$-matrix shown by eq.~(\ref{p1.18}) the permutation $P$
takes the form
\begin{eqnarray}
P=\left[\begin{array}{cccc}1& & &\\ &0&1& \\ &1&0& \\ & & &1\end{array}
\right]\label{p2.1}
\end{eqnarray}
Setting the transfer matrix
\begin{eqnarray}
T(u)=\left[\begin{array}{cc}T_{11}(u)&T_{12}(u)\\T_{21}(u)&T_{22}(u)
\end{array}\right]=\sum^{\infty}_{n=0}u^{-n}T^{(n)}\label{p2.2}
\end{eqnarray}
where
\begin{equation}
T_{ab}(u)=\sum^{\infty}_{n=0}u^{-n}T^{(n)}_{ab}\ \ (a,b=1,2)\label{p2.3}
\end{equation}
and substituting eqs.~(\ref{p1.18}) and (\ref{p2.2}) into eq.~(\ref{p1.7})
we find the relations satisfied by $T_{ab}(u)\ (a,b,c,d=1,2)$:
\begin{equation}
(u-v)[T_{bc}(u),T_{ad}(v)]+T_{ac}(u)T_{bd}(v)-T_{ac}(v)T_{bd}(u)=0\label{p2.4}
\end{equation}
that can be expanded in terms of $u$ and $v$ through eq.(\ref{p2.3})
($m,n=0,1,\ldots,\infty$)
\begin{eqnarray}
\lefteqn{[T^{(0)}_{ab}, T^{(n)}]=0\label{p2.5}}\\
\lefteqn{[T^{(n+1)}_{bc},T^{(m)}_{ad}]-[T^{(n)}_{bc},T^{(m+1)}_{ad}]
+T^{(n)}_{ac}T^{(m)}_{bd}-T^{(m)}_{ac}T^{(n)}_{bd}=0\label{p2.6}}
\end{eqnarray}
eqs.~(\ref{p2.5}) and (\ref{p2.6}) have the following independent relations
($a,b,c,d=1,2$)
\begin{eqnarray}
\lefteqn{\hspace*{0.55cm}[T^{(m)}_{ab}, T^{(n)}_{ab}]=0,\ \
[T^{(m)}_{ab},T^{(n)}_{cd}]=
[T^{(n)}_{ab},T^{(m)}_{cd}]  \label{p2.7} }\\
\lefteqn{\left\{\begin{array}{l}
{[T^{(n+1)}_{11},T^{(m)}_{12}]}-[T^{(n)}_{11},T^{(m+1)}_{12}]
+T^{(n)}_{11}T^{(m)}_{12}-T^{(m)}_{11}T^{(n)}_{12}=0\\
{[T^{(n+1)}_{11},T^{(m)}_{21}]}-[T^{(n)}_{11},T^{(m+1)}_{21}]
+T^{(n)}_{21}T^{(m)}_{11}-T^{(m)}_{21}T^{(n)}_{11}=0\\
{[T^{(n+1)}_{22},T^{(m)}_{12}]}-[T^{(n)}_{22},T^{(m+1)}_{12}]
+T^{(n)}_{12}T^{(m)}_{22}-T^{(m)}_{12}T^{(n)}_{22}=0\\
{[T^{(n+1)}_{22},T^{(m)}_{21}]}-[T^{(n)}_{22},T^{(m+1)}_{21}]
+T^{(n)}_{22}T^{(m)}_{21}-T^{(m)}_{22}T^{(n)}_{21}=0
\end{array}\label{p2.8}\right.}\\
\lefteqn{\left\{\begin{array}{l}
{[T^{(n+1)}_{11},T^{(m)}_{22}]}-[T^{(n)}_{11},T^{(m+1)}_{22}]
+T^{(n)}_{21}T^{(m)}_{12}-T^{(m)}_{21}T^{(n)}_{12}=0\\
{[T^{(n+1)}_{12},T^{(m)}_{21}]}-[T^{(n)}_{12},T^{(m+1)}_{21}]
+T^{(n)}_{22}T^{(m)}_{11}-T^{(m)}_{22}T^{(n)}_{11}=0
\end{array}\label{p2.9}\right.}
\end{eqnarray}
Obviously, $T^{(0)}_{ab}$ must be a $c$-numbered matrix. Because the form of
RTT relation is invariant under a similar transformation, without loss of
generality two possible forms of matrix $T^{(0)}$
\begin{eqnarray}
(\mbox{A}) \ \ \ T^{(0)}=\left[\begin{array}{cc}1&0\\0&\mu\end{array}\right]\ \
(\mu\neq 0) \label{p2.10}
\end{eqnarray}
and
\begin{eqnarray}
(\mbox{B}) \ \ \ T^{(0)}=\left[\begin{array}{cc}1&0\\0&0\end{array}
\right]\ \ \label{p2.11}
\end{eqnarray}
are allowed.

\subsection{$Y(sl(2))$ Symmetry}

In this section only (A) is taken into account. In this case
eqs.~(\ref{p2.7})-(\ref{p2.9}) are reduced to the following independent set
of relations:
\begin{eqnarray}
\lefteqn{\hspace*{0.55cm}[T^{(m)}_{ab}, T^{(n)}_{ab}]=0,\ \
[T^{(m)}_{ab},T^{(n)}_{cd}]=
[T^{(n)}_{ab},T^{(m)}_{cd}]\ (n,m\geq 1)  \label{p2.09} }\\
\lefteqn{\left\{\begin{array}{ll}
{[T^{(n)}_3(\mu),T^{(1)}_{12}]}=2\mu T^{(n)}_{12}&
[T^{(n)}_{12},T^{(1)}_{21}]=T^{(n)}_3(\mu)\\
{[T^{(n)}_3(\mu),T^{(1)}_{21}]}=-2\mu T^{(n)}_{21}&
[T^{(n)}_0(\mu),T^{(1)}_{ab}]=0\ (n=1,2,\ a,b=1,2)
\end{array}\label{p2.12}\right.}
\end{eqnarray}
where
\begin{eqnarray}
\lefteqn{\left\{\begin{array}{ll} T^{(n)}_3(\mu)=T^{(n)}_{22}-\mu
T^{(n)}_{11},&
T^{(n)}_0(\mu)=T^{(n)}_{22}+\mu T^{(n)}_{11},\\ T^{(n)}_+=T^{(n)}_{12},
&T^{(n)}_-=T^{(n)}_{21}\end{array}\right.\label{p2.13}}\\
\lefteqn{\left\{\begin{array}{l} [T^{(2)}_0(\mu), T^{(2)}_3(\mu)]+2\mu
(T^{(1)}_{21}
T^{(2)}_{12}-T^{(1)}_{21}T^{(1)}_{12})=0\\
{[T^{(n)}_0(\mu), T^{(2)}_{12}]}+T^{(1)}_{12}
T^{(n)}_3(\mu)-T^{(n)}_{12}T^{(1)}_3(\mu)=0\ \ (n\geq 1)\\
{[T^{(n)}_0(\mu), T^{(2)}_{21}]}+T^{(n)}_{21}T^{(1)}_3(\mu)-T^{(1)}_{21}
T^{(n)}_3(\mu)=0\ \ (n\geq 1)\end{array}\right.\label{p2.14}}
\end{eqnarray}
and
\begin{eqnarray}
\lefteqn{T^{(n+1)}_{12}=(2\mu)^{-1}\{[T^{(n)}_3(\mu), T^{(2)}_{12}]+
T^{(1)}_{12}T^{(n)}_0(\mu)-T^{(n)}_{12}T^{(1)}_0(\mu)\}\nonumber}\\
\lefteqn{T^{(n+1)}_{21}=(2\mu)^{-1}\{[T^{(2)}_{21},
T^{(n)}_3(\mu)]+T^{(1)}_{21}
T^{(n)}_0(\mu)-T^{(n)}_{12}T^{(1)}_0(\mu)\}\label{p2.15}}\\
\lefteqn{T^{(n+1)}_3(\mu)=[T^{(n)}_{12}, T^{(2)}_{21}]+T^{(1)}_{22}
T^{(n)}_{11}-T^{(n)}_{22}T^{(1)}_{11},\ (n\geq 2)\nonumber}
\end{eqnarray}
It is emphasized that because of the iterative relation eq.~(\ref{p2.15}),
only $T^{(1)}$ and $T^{(2)}$ are basic ones. To satisfy all the relations with
$T^{(n)}(n\geq 3)$ it is enough to look for the constraints to $T^{(3)}$
that in turn to provide the constraints to $T^{(2)}$ itself. Before doing
this let us introduce the ``physical'' operators:
\begin{eqnarray}
\lefteqn{T^{(1)}_{12}=\alpha_+I_+,\ \ T^{(1)}_{21}=\alpha_-I_-,\ \
T^{(1)}_3(\mu)=2\mu I_3,}\label{p2.16}\\
\lefteqn{T^{(2)}_{12}=\beta_+J_+,\ \ T^{(2)}_{21}=\beta_-J_-,\ \
T^{(2)}_3(\mu)=2\beta_3 J_3,}\label{p2.17}
\end{eqnarray}
where $\alpha_{\pm},\ \beta_{\pm}$ and $\beta_3$ are constants to be
determined.

Substituting eqs.~(\ref{p2.14}) and (\ref{p2.15}) into eqs.~(\ref{p2.09})
and (\ref{p2.12}) one obtains
\begin{equation}
\alpha_+\alpha_-=\mu, \ \ \alpha_+\beta_-=\alpha_-\beta_+=\beta_3\label{p2.18}
\end{equation}
and the algebraic relations
\begin{eqnarray}
\lefteqn{[I_3,I_{\pm}]=\pm I_{\pm},\ \ \ [I_+,I_-]=2I_3\nonumber}\\
\lefteqn{[I_3,J_{\pm}]=[J_3,I_{\pm}]=\pm J_{\pm},\ \ [I_{\pm},J_{\mp}]
=\pm 2J_3} \label{p2.19}\\
\lefteqn{[I_{\pm},J_{\pm}]=[I_3,J_3]=0}\nonumber
\end{eqnarray}
or by
\begin{equation}
I_{\pm}=I_1\pm i I_2,\ \ \ J_{\pm}=J_1\pm i J_2 \label{p2.20}
\end{equation}
we have
\begin{eqnarray}
[I_{\lambda},I_{\mu}]&=&C_{\lambda\mu\nu}I_{\nu},\ \
[I_{\lambda},J_{\mu}]=C_{\lambda\mu\nu}J_{\nu},\label{p2.21}\\
C_{\lambda\mu\nu}&=&i\epsilon_{\lambda\mu\nu},\ (\lambda,\mu,\nu=1,2,3)
\label{p2.39}
\end{eqnarray}
Next we have to look for the constraints to $T^{(3)}$ by taking
eqs.~(\ref{p2.11})-(\ref{p2.15}) into account. It is easy to know from
eq.~(\ref{p2.09}) that
\begin{equation}
[T^{(2)}_{22}, T^{(1)}_{11}]=[T^{(3)}_{22}, T^{(2)}_{11}]=0\label{p2.021}
\end{equation}
{}From eq.~(\ref{p2.15}) it follows
\begin{equation}
T^{(3)}_3(\mu) \equiv T^{(3)}_{22}-\mu
T^{(3)}_{11}=[T^{(2)}_{12},T^{(2)}_{21}]+T^{(1)}_{22}
T^{(2)}_{11}-T^{(2)}_{22}T^{(1)}_{11}\label{p2.22}
\end{equation}
By taking the commutator between $T^{(2)}_{22}$ and both of sides of
eq.~(\ref{p2.22}) and using eq.~(\ref{p2.021}) we obtain
\begin{equation}
\mu[T^{(3)}_{11},T^{(2)}_{22}]=[T^{(2)}_{22},[T^{(2)}_{12},T^{(2)}_{21}]]
+T^{(1)}_{22}[T^{(2)}_{22},T^{(2)}_{11}]\label{p2.23}
\end{equation}
Similarly
\begin{equation}
[T^{(2)}_{11},T^{(3)}_{22}]=[T^{(2)}_{11},[T^{(2)}_{12},T^{(2)}_{21}]]
+[T^{(2)}_{22},T^{(2)}_{11}]T^{(1)}_{11}\label{p2.24}
\end{equation}
Hence, the relation eq.~(\ref{p2.9}) for $[T^{(3)}_{11},T^{(2)}_{22}]=
[T^{(2)}_{11},T^{(3)}_{22}]$ leads to
\begin{equation}
[T^{(2)}_3(\mu),[T^{(2)}_{12},T^{(2)}_{21}]]=T^{(1)}_3(\mu)(T^{(1)}_{21}T^{(2)}_{12}
-T^{(2)}_{21}T^{(1)}_{12}).\label{p2.25}
\end{equation}
With the same manipulations to $T^{(3)}_{12}$ in eq.~(\ref{p2.15}) one gets
\begin{eqnarray}
\lefteqn{[T^{(2)}_{12},T^{(3)}_{12}]=[T^{(2)}_{12},[T^{(2)}_{12},T^{(2)}_{11}]]
+T^{(1)}_{12}[T^{(2)}_{12},T^{(2)}_{11}]-T^{(2)}_{12}
[T^{(2)}_{12},T^{(2)}_{11}]\label{p2.26}}\\
\hspace*{-0.25cm}\lefteqn{\mu[T^{(3)}_{12},T^{(2)}_{12}]=[T^{(2)}_{12},[T^{(2)}_{22},T^{(2)}_{12}]]
+T^{(1)}_{12}[T^{(2)}_{12},T^{(2)}_{22}]-T^{(2)}_{12}
[T^{(2)}_{12},T^{(2)}_{22}]\label{p2.27}}
\end{eqnarray}
{}From $[T^{(2)}_{12}, T^{(3)}_{12}]=0$ it follows
\begin{equation}
[T^{(2)}_{12},[T^{(2)}_3(\mu),
T^{(2)}_{12}]]=T^{(1)}_{12}(T^{(2)}_{12}T^{(1)}_3(\mu)
-T^{(1)}_{12}T^{(2)}_3(\mu))\label{p2.28}
\end{equation}
Similarly the relations $[T^{(2)}_{21}, T^{(3)}_{21}]=0$ provides the
constraint
\begin{equation}
[T^{(2)}_{21},[T^{(2)}_3(\mu),
T^{(2)}_{21}]]=T^{(1)}_{21}(T^{(2)}_{21}T^{(1)}_3(\mu)
-T^{(1)}_{21}T^{(2)}_3(\mu))\label{p2.29}
\end{equation}
By taking $[T^{(2)}_{12}, T^{(3)}_3(\mu)]$ and $[T^{(2)}_{21}, T^{(3)}_3(\mu)]$
where
$T^{(3)}_3(\mu)$ is given by eq.~(\ref{p2.15}) and considering eq.~(\ref{p2.9})
we obtain
\begin{eqnarray}
\lefteqn{\hspace*{0.4cm}2\mu[T^{(2)}_{12},[T^{(2)}_{12}, T^{(2)}_{21}]]+
[T^{(2)}_3(\mu), [T^{(2)}_3(\mu),T^{(2)}_{12}]]}\nonumber\\
\lefteqn{=(T^{(2)}_{12}T^{(3)}_3(\mu)-T^{(1)}_{12}T^{(2)}_3(\mu))(T^{(1)}_3
(\mu)+2\mu)\nonumber}\\
\lefteqn{\hspace*{0.4cm}-2\mu
T^{(1)}_{12}(T^{(2)}_{21}T^{(1)}_{12}-T^{(1)}_{21}
T^{(2)}_{12})\label{p2.30}}
\end{eqnarray}
and
\begin{eqnarray}
\lefteqn{\hspace*{0.4cm}2\mu[T^{(2)}_{21},[T^{(2)}_{12}, T^{(2)}_{21}]]+
[T^{(2)}_3(\mu),[T^{(2)}_{21},T^{(2)}_3(\mu)]]\nonumber}\\
\lefteqn{=(T^{(2)}_{21}T^{(3)}_3(\mu)-T^{(1)}_{21}T^{(2)}_3(\mu))(T^{(1)}_3
(\mu)-2\mu)\nonumber}\\
\lefteqn{\hspace*{0.4cm}+2\mu
T^{(1)}_{21}(T^{(2)}_{21}T^{(1)}_{12}-T^{(1)}_{21}
T^{(2)}_{12})\label{p2.31}}
\end{eqnarray}
we thus have exhausted all the relations including $T^{(3)}$ in
eqs.~(\ref{p2.9})-(\ref{p2.15}). Substituting eqs.~(\ref{p2.16}) and
(\ref{p2.17}) into eqs.~(\ref{p2.25})-(\ref{p2.31}) we derive
\begin{eqnarray}
\lefteqn{[J_3,[J_+,J_-]]=\frac{\mu}{\beta_+\beta_-}I_3(J_-I_+-I_-J_+)\label{p2.32}}\\
\lefteqn{{[J_{\pm},[J_3,J_{\pm}]]}=\frac{\mu}{\beta_+\beta_-}I_{\pm}
(J_{\pm}I_3-I_{\pm}J_3)\label{p2.33}}\\
\lefteqn{2[J_{3},[J_3,J_{\pm}]]+[J_{\pm},[J_{\pm},J_{\mp}]]}\nonumber\\
\lefteqn{=\frac{\mu}{\beta_+\beta_-}\{2I_{3}(J_{\pm}I_3-I_{\pm}J_3)
+I_{\pm}(I_{-}I_+-J_{-}I_+)\}\label{p2.34}}
\end{eqnarray}
By introducing
\begin{equation}
\mu\beta^{-1}_+\beta^{-1}_-=\frac 1 4 h^2 \label{p2.35}
\end{equation}
the eqs.~(\ref{p2.32})-(\ref{p2.34}) can be recast into
\begin{eqnarray}
&&[[J_{\lambda},J_{\mu}],[I_{\tau},J_{\rho}]]+[[J_{\tau},J_{\rho}],
[I_{\lambda},J_{\mu}]]\nonumber\\
&=&h^2(a_{\lambda\mu\nu\alpha\beta\gamma}C_{\tau\rho\nu}+a_{\tau\rho\nu
\alpha\beta\gamma}C_{\lambda\mu\nu})\{I_{\alpha},I_{\beta},I_{\gamma}\}
\label{p2.36}
\end{eqnarray}
where
\begin{eqnarray}
a_{\lambda\mu\nu\alpha\beta\gamma}&=&\frac1{24}C_{\lambda\rho\sigma}
C_{\mu\beta\rho}C_{\nu\gamma\tau}C_{\sigma\gamma\tau}\label{p2.37}\\
\{x_1,x_2,x_3\}&=&\sum_{i\neq j\neq k \neq i}x_ix_jx_k\label{p2.38}
\end{eqnarray}
With eq.~(\ref{p2.39}) it is easy to prove that
\begin{equation}
[J_{\lambda},[J_{\mu},I_{\nu}]]-[I_{\lambda},[J_{\mu},J_{\nu}]]=0\label{p2.40}
\end{equation}

In conclusion for given $R$-matrix eq.~(\ref{p1.18}) by solving RTT relation
we find the infinite algebra shown by eqs.~(\ref{p2.21}), (\ref{p2.36})
and (\ref{p2.40}). This algebra is not closed since $J_{\mu}$ are not closed.
Moreover the coproduct $\Delta$ is defined by
\begin{equation}
\Delta T_{ab}(u)=\sum_cT_{ac}(u)\otimes T_{cb}(u)
\end{equation}
that by setting $T^{(1)}_{22}=\mu I_3$ and $T^{(1)}_{11}=-\mu I_3$ leads to
\begin{eqnarray}
\Delta(I_{\lambda})&=&I_{\lambda}\otimes 1+1\otimes I_{\lambda}\label{pp2.43}\\
\Delta(J_{\lambda})&=&J_{\lambda}\otimes 1+1\otimes J_{\lambda}
+\frac 12 hC_{\lambda\mu\nu}I_{\nu}\otimes I_{\mu}\label{pp2.44}.
\end{eqnarray}
This set of infinite algebra is called $Y(sl(2))$.

\subsection{Drinfeld's Theory of {\em Yangian}}

Let $g$ be a finite-dimensionally simple Lie algebra the associated algebra
$A$ is generated by elements $I_{\lambda}$ and $J_{\lambda}$ with defining
relations
\begin{eqnarray}
\lefteqn{[I_{\lambda}, I_{\mu}]=C_{\lambda\mu\nu}I_{\nu}, \ \
[I_{\lambda}, J_{\mu}]=C_{\lambda\mu\nu}J_{\nu},\label{p2.41}}\\
\lefteqn{[J_{\lambda},[J_{\mu},I_{\nu}]]-[I_{\lambda},[J_{\mu},I_{\nu}]]
=h^2a_{\lambda\mu\nu\alpha\beta\gamma}\{I_{\alpha},I_{\beta},I_{\gamma}\}
\label{p2.42}}\\
\lefteqn{[[J_{\lambda},J_{\mu}],[I_{r},J_{s}]]+[[J_{r},J_{s}],
[I_{\lambda},J_{\mu}]]\nonumber}\\
\lefteqn{=h^2(a_{\lambda\mu\nu\alpha\beta\gamma}C_{rs\nu}+a_{rs\nu\alpha\beta
\gamma}C_{\lambda\mu\nu})\{I_{\alpha},I_{\beta},I_{\gamma}\}\label{p2.43}}
\end{eqnarray}
where the $C_{\lambda\mu\nu}$ are the structure constants of $g$,
$a_{\lambda\mu\nu\alpha\beta\gamma},\ \{x_1,x_2,x_3\}$ are given in
eqs.~(\ref{p2.37}) and (\ref{p2.38}) and all the repeating indices
imply summation. Moreover, the co-multiplication is given by
\begin{eqnarray}
\Delta(I_{\lambda})&=&I_{\lambda}\otimes 1+1\otimes I_{\lambda}\label{pp2.46}\\
\Delta(J_{\lambda})&=&J_{\lambda}\otimes 1+1\otimes J_{\lambda}
+\frac 12 h C_{\lambda\mu\nu}I_{\nu}\otimes I_{\mu}\label{p2.46}
\end{eqnarray}
that are the same as eqs.~(\ref{pp2.43}) and (\ref{pp2.44}) for $sl(2)$.
Set eqs.~(\ref{p2.41})-(\ref{p2.46}) is called {\em Yangian} denoted by $Y(g)$.
Setting $h=1$ it is a Hopf algebra. When $g=sl(2)$ eq.~(\ref{p2.42}),
i.e., eq.~(\ref{p2.40}) is identically zero. However eq.~(\ref{p2.43}) holds
that has been shown in the above calculation. For $g=sl(n)(n>2)$ Drinfeld
pointed out that~\cite{re10} eq.~(\ref{p2.43}) follows from eqs.~(\ref{p2.41})
and (\ref{p2.42}) which is usually called serre relations(There is a mis-print
in the remark appeared in p255 of Ref.~\cite{re10}).

On one word for $sl(2)$ eqs.~(\ref{p2.41}) and (\ref{p2.43}) should only
be concerned, whereas for $sl(n)(n>2)$ only eqs.~(\ref{p2.41}) and
(\ref{p2.42}) should be taken into account. In the \S2.1 the
$Y(sl(2))$ has explicitly been obtained based on the RTT relation. let
us take $Y(sl(2))$ as an example to show the properties of {\em Yangian}.
\begin{enumerate}
\item It contains the classical algebra $sl(2)$ as a sub-algebra denoted
by the set $\{I_{\lambda}\}(\lambda=1,2,3)$ that is closed.
\item Besides the set $\{I_{\lambda}\}$ there appears a new set $\{J_{\lambda}
\}(\lambda=1,2,3)$ which are not closed. Hence with $\{I_{\lambda}\}$  and
$\{J_{\lambda}\}$ an infinite algebra can be generated.
\item Observing the expansion eq.~(\ref{p2.3}) it looks like a loop algebra
with the deference that the $T^{(n)}_{ab}$ obey the ``deformed'' commutation
relations. Thus we may regard {\em Yangian} as a deformed loop algebra.
We have known that both of sides of eq.~(\ref{p2.42}) vanish identically
for $sl(2)$. The spectral parameter $u$ plays very important role in the
{\em Yangian} theory. Just the spectral parameter dependence gives rise
to the new operators $\{J_{\lambda}\}$ beyond the usual Lie algebra $sl(2)$,
and make the infinite algebra not the loop algebra.
\item The RTT relation provides a standard method to generate {\em Yangian}
even
though the calculations are tedious. From the RTT relation we have derived the
commutation relations satisfied by $\{I_{\lambda}\}$ and $\{J_{\lambda}\}$
that are quantum operators. The realizations of such operators can be made
through the usual quantum mechanical operators or second quantized operations.
The {\em Yangian} symmetry is model-independent. It is related to a given Lie
algebra. However a realization is model-dependent. Hence it is connected
with the specified physical model.
\end{enumerate}

\subsection{$\mbox{det}T(u)$ Associated with $Y(gl(2))$}

For $sl(2)$ a $R$-matrix is given by eq.~(\ref{p1.18}) and the $T(u)$ is 2 by 2
matrix. Each elements of $T(u)$ matrix is quantum operator. The inverse of
$T(u)$ is defined by
\begin{eqnarray}
T^{-1}(u)=\frac 1{\mbox{det}T(u)}\left[\begin{array}{cc}\tilde{T}_{11}(u)
&\tilde{T}_{12}(u)\\ \tilde{T}_{21}(u)&\tilde{T}_{22}(u)\end{array}
\right]\label{p2.47}
\end{eqnarray}
and $\mbox{det}T(u)$ should commute with $T_{ab}(u)(a,b=1,2)$:
\begin{equation}
[\mbox{det}T(u), T_{ab}(v)]=0,\ (a,b=1,2),\label{p2.48}
\end{equation}
i.e., $\mbox{det}T(u)$ is a center of the algebra. From eq.~(\ref{p2.4}) by
setting
$u-v=1$, it follows
\begin{equation}
[T_{bc}(u), T_{ad}(u-1)]-T_{ac}(u)T_{bd}(u-1)-T_{ac}(u-1)T_{bd}(u)=0
\label{p2.49}
\end{equation}
Substituting eq.~(\ref{p2.47}) into $T^{-1}(u)T(u)=T(u)T^{-1}(u)=1$ and making
comparison to eq.~(\ref{p2.49}) we have
\begin{eqnarray}
\begin{array}{ll}\tilde{T}_{12}(u)=-{T}_{12}(u-1)& \tilde{T}_{21}(u)
=-{T}_{12}(u-1)\\ \tilde{T}_{11}(u)={T}_{22}(u-1)&\tilde{T}_{22}(u)
={T}_{11}(u-1)\end{array}\label{p2.50}
\end{eqnarray}
and~\cite{re10}
\begin{equation}
\mbox{det}T(u)=T_{11}(u)T_{22}(u-1)-T_{12}(u)T_{21}(u-1).\label{p2.51}
\end{equation}
The direct check also verifies the validity of eq.~(\ref{p2.48}) by virtue
of eq.~(\ref{p2.51}). Of course, it holds
\begin{equation}
[\mbox{det}T(u),\mbox{tr}T(v)]=0,\ (a,b=1,2),\label{p2.52}
\end{equation}
The conserved family  $C_n$ determined by
\begin{equation}
\mbox{det}T(u)=\sum^{\infty}_{n=0}u^{-n}C_n\label{p2.53}
\end{equation}
can be calculated explicitly:
\begin{eqnarray}
C_0&=&1,\ \ C_1=T^{(1)}_{11}+\mu T^{(1)}_{22}=T^{(1)}_0(\mu)\nonumber\\
C_j &=& T^{(j)}_{0}(\mu)+ \sum_{\scriptsize \begin{array}{c}m+l=j\\m,l\neq 0
 \end{array}} \frac{(l+m-1)!}{(m-1)!l!}T^{(m)}_{22}+\nonumber\\
 &&\sum_{\scriptsize \begin{array}{c}m+n+l=j\\m,n\neq 0 \end{array}}
\frac{(l+m-1)!}
 {(m-1)!l!}\left( T^{(n)}_{11}T^{(m)}_{22}-T^{(n)}_{12}T^{(m)}_{21}\right)
 \ (j\geq 1)\label{p2.54}.
\end{eqnarray}
The explicit forms of $C_k$ are:
\begin{eqnarray}
C_1&=&T^{(1)}_{22}+\mu T^{(1)}_{11}=T^{(1)}_0(\mu)=\mu N,\label{p2.55}\\
C_2&=&T^{(2)}(\mu)+T^{(1)}_{22}+T^{(1)}_{11}T^{(1)}_{22}-T^{(1)}_{12}
T^{(1)}_{21}=\mu(M+\frac 12 N+\frac 12 N^2)\label{p2.56}\\
C_3&=&T^{(3)}_0(\mu)+T^{(2)}_3(\mu)+T^{(2)}_{11}T^{(1)}_{22}-T^{(2)}_{12}
T^{(1)}_{21}+T^{(1)}_{11}T^{(2)}_{22}-T^{(1)}_{12}T^{(2)}_{21}+C_2\nonumber\\
&=&\frac 12 \mu MN\label{p2.57}\\
& &\cdots     \nonumber
\end{eqnarray}
where
\begin{eqnarray}
T^{(1)}_0(\mu)&\equiv& \mu N, \label{p2.58}\\
T^{(2)}_0(\mu)&\equiv& T^{(2)}_{22}+\mu T^{(2)}_{11}=\mu(\vec{I}^2+M),
\label{p2.59}
\end{eqnarray}
with
\begin{equation}
\vec{I}^2=\sum^3_{\lambda=1} I^2_{\lambda}\label{p2.60}
\end{equation}
the casimir of $gl(2)$ algebra.

It can easily be proved that
\begin{equation}
[C_m, J^{(n)}]=0,\ \ \ [C_m,C_n]=0\label{p2.61}
\end{equation}
for any $\mu$. Therefore $C_m$ commute with any conserved quantities.

Finally, let us summarize the RTT relation shown by eqs.~(\ref{p2.5}) and
(\ref{p2.6}) that can be reduced to the following independent set of relations
$(T^{(n)}_3=T^{(n)}_3(\mu),\ T^{(n)}_0=T^{(n)}_0(\mu),\
T^{(n)}_+=T^{(n)}_{12},\ T^{(n)}_-=T^{(n)}_{21})$:
\begin{eqnarray}
\lefteqn{\left\{\begin{array}{ll}
[T^{(1)}_{\alpha},T^{(2)}_{\alpha}]=0,\ (\alpha=\pm, 3)\\
{[T^{(1)}_{3},T^{(k)}_{\pm}]}=[T^{(k)}_{3},T^{(1)}_{\pm}]=\pm 2 \mu
T^{(k)}_{\pm},\ (k=1,2)\\
{[T^{(1)}_{+},T^{(k)}_{-}]}=[T^{(k)}_{+},T^{(1)}_{-}]=T^{(k)}_{3},
\end{array}\right.}\label{p2.62}\\
\lefteqn{\left\{\begin{array}{ll}
[T^{(n)}_{0},T^{(m)}_{0}]=0,\ (\forall n,m)\\
{[T^{(1)}_{0},T^{(n)}_{\alpha}]}=[T^{(n)}_{0},T^{(1)}_{\alpha}]=0,\
(\alpha=\pm,\ 3),\\
{[T^{(2)}_{0},T^{(2)}_{\pm}]}=\pm(T^{(1)}_{3}T^{(2)}_{\pm}-T^{(2)}_3
T^{(1)}_{\pm}),\\
{[T^{(2)}_{0},T^{(2)}_{3}]}=2\mu (T^{(1)}_{+}T^{(2)}_{-}-T^{(2)}_{+}T^{(1)}_-),
\end{array}\right.}\label{p2.63}\\
\lefteqn{\left\{\begin{array}{ll}
T^{(n+1)}_{\pm}=(2\mu)^{-1}\{\pm [T^{(2)}_{3},T^{(n)}_{\pm}]+T^{(n)}_{0}
T^{(1)}_{\pm}-T^{(1)}_{0}T^{(n)}_{\pm}\}\\
T^{(n+1)}_{3}= [T^{(n)}_{+},T^{(2)}_{-}]+(2\mu)^{-1}(T^{(n)}_{0}T^{(1)}_{3}
-T^{(1)}_{0}T^{(n)}_{3})
\end{array}\right.}\label{p2.64}
\end{eqnarray}
and
\begin{equation}
[T^{(n)}_{\alpha},T^{(m)}_{\alpha}]=0\ (\alpha\neq 0,\ m<n,\ n>2)\label{p2.65}
\end{equation}
which is equivalent to
\begin{eqnarray}
\left\{\begin{array}{ll}
[T^{(m)}_{\pm},[T^{(2)}_3, T^{(n)}_{\pm}]]\pm [T^{(m)}_{\pm},T^{(n)}_{0}]
T^{(1)}_{\pm}=0\\
2\mu[T^{(m)}_{3},[T^{(n)}_+, T^{(2)}_{-}]]+
[T^{(m)}_{3},T^{(n)}_{0}]T^{(1)}_3=0
\end{array}\right.\label{p2.66}
\end{eqnarray}
To show the relationship for the set of eqs.~(\ref{p2.62})-(\ref{p2.65}) and
the set of eqs.~(\ref{p2.21}), (\ref{p2.40}) and (\ref{p2.36}) i.e., the
Drinfeld form for $Y(sl(2))$, by taking $m=n=2$ in eqs.~(\ref{p2.62})
and (\ref{p2.66}) it follows
\begin{eqnarray}
\lefteqn{\hspace*{0.4cm}2\mu[T^{(2)}_{\pm},[T^{(2)}_+,T^{(2)}_-]]\pm
[T^{(2)}_3,[T^{(2)}_3,T^{(2)}_{\pm}]]\nonumber}\\
\lefteqn{=2\mu(T^{(1)}_+T^{(2)}_--T^{(2)}_+T^{(1)}_-)T^{(1)}_{\pm}\pm
(T^{(1)}_3T^{(2)}_{\pm}-T^{(2)}_3T^{(1)}_{\pm})T^{(1)}_3}\label{p2.67}
\end{eqnarray}
Eqs.~(\ref{p2.62}), (\ref{p2.28}) and (\ref{p2.31}) lead to
eqs.~(\ref{p2.21}), (\ref{p2.36}) and (\ref{p2.40}), i.e., $Y(sl(2))$
can be generated based on $T^{(1)}$ and $T^{(2)}$. The ``closure'' of
eqs.~(\ref{p2.21}), (\ref{p2.36}) and (\ref{p2.40}) for $sl(2)$ means
that we can use $\{I_{\lambda}\}$ and $\{J_{\lambda}\}$ to generate any
new generators of $Y(sl(2))$ based on eq.~(\ref{p2.64}) to form an
infinite algebra.

\section{Realization of {\em Yangian}}
\setcounter{equation}{0}

\subsection{$Y(sl(2))$}
\begin{equation}
\vec{I}=\sum^N_{i=1}\vec{S}_i,\ \ \vec{J}=\sum^N_{i\neq j}\vec{S}_i\times
\vec{S}_j \label{p3.1}
\end{equation}
where $\vec{I}=\{I_{\lambda}\}$ and $\vec{J}=\{J_{\lambda}\}(\lambda=1,2,3)$
and
\begin{equation}
[S^{\lambda}_i, S^{\mu}_j]
=i\epsilon_{\lambda\mu\nu}S^{\nu}_i\delta_{ij}\label{p3.2}
\end{equation}
satisfy eqs.~(\ref{p2.41})-(\ref{p2.43}). Therefore eq.~(\ref{p3.1}) forms
a $Y(sl(2))$~\cite{re11}.

To prove the statement first we rewrite eq.~({\ref{p3.2})
\begin{eqnarray}
\lefteqn{[S^{\alpha}_i,S^{\alpha}_j]=0,\ \ \mbox{for}\ \alpha=\pm,\
3,\nonumber}\\
\lefteqn{[S^{3}_i,S^{\pm}_j]=\pm\delta_{ij}S^{\pm}_i,\ \
[S^{+}_i,S^-_j]=2\delta_{ij}S^{3}_i\nonumber}
\end{eqnarray}
We should verify that the defined $\{I_{\lambda}\}$ and $\{J_{\lambda}\}$
satisfy eq.~(\ref{p2.19}) and eqs.~(\ref{p2.32})-(\ref{p2.34}). Since by
$J_{\lambda}\rightarrow h^{\frac 12}J_{\lambda}$ the constant $h$ can be
removed, one can takes $h=1$.

Defining
\begin{eqnarray}
\lefteqn{I_{\pm}=\sum_iS^{\pm}_i,\hspace{1cm} I_3=\sum_iS^3_i,\nonumber}\\
\lefteqn{J_{\pm}=\mp \sum_{i\neq j}W_{ij}S^{\pm}_iS^3_j,\ \ J_3=\frac 12
\sum_{i\neq j} W_{ij}S^+_iS^-_j}
\end{eqnarray}
where $W_{ij}+W_{ji}=0$, and substituting them into eqs.~(\ref{p2.19}),
(\ref{p2.32})-(\ref{p2.34}) we have
\[(S^{\pm}_i)^2=0,\ \ (S^3_i)^2=1 \]
i.e., the operators $S^{\lambda}_i$ should be Pauly matrices, and
\begin{equation}
W_{ij}W_{jk}+W_{jk}W_{ki}+W_{ki}W_{ij}=-h^2,\ (\mbox{for}\ i\neq j\neq k\neq i)
\label{pw}]
\end{equation}
The simplest solution of $W_{ij}$ is given by
\begin{eqnarray}
W_{ij}=\left\{\begin{array}{cl}h&i<j\\0&i=j\\-h&i>j\end{array}\right.\label{pw1}
\end{eqnarray}
that is nothing but the statement.

\subsection{Example for $Y(sl(2))$}

By using the fermion operators acting on i-th site
\begin{equation}
a^{\dag}_i=c^{\dag}_{i\uparrow}, \ \ a_i=c_{i\uparrow}, \ \
b^{\dag}_i=c^{\dag}_{i\downarrow},\ \ b_i=c_{i\downarrow}\label{p3.3}
\end{equation}
satisfying
\begin{eqnarray}
\lefteqn{\{a_i,a_j\}=\{a^{\dag},a^{\dag}_j\}=\{b_i,b_j\}=\{b^{\dag}_i,
b^{\dag}_j\}=0\label{p3.4}}\\
\lefteqn{\{a^{\dag}_i,a_j\}=\delta_{ij}, \ \ \ \{b^{\dag}_i,b_j\}
=\delta_{ij}, \label{p3.5}}
\end{eqnarray}
and $a^{\dag}_i(a_i)$ anticommute with $b^{\dag}_i(b_i)$. The realization of
$Y(sl(2))$ can be made by
\begin{eqnarray}
I_+ &=& \sum_i a^{\dag}_ib_i=\sum_iI^+_i\nonumber\\
I_- &=& \sum_i b^{\dag}_ia_i=\sum_iI^-_i\label{p3.6}\\
I_3 &=& \frac 12\sum_i(a^{\dag}_ia_i-b^{\dag}_ib_i)=\sum_iI^3_i\nonumber
\end{eqnarray}
that satisfy $sl(2)$ algebra
\begin{equation}
[I_3, I_{\pm}]=\pm I_{\pm},\ \ [I_+,I_-]=2I_3\label{p3.7}
\end{equation}
and
\begin{eqnarray}
J_+&=&\sum_{i,j}\theta_{i,j}a^{\dag}_ib_j-U\sum_{i,j}\epsilon_{i,j}I^+_iI^3_j
=\sum_iJ^+_i\nonumber\\
J_-&=&\sum_{i,j}\theta_{i,j}b^{\dag}_ia_j+\sum_{i,j}\epsilon_{i,j}I^-_iI^3_j
=\sum_iJ^-_i\label{p3.8}\\
J_3&=&\frac 12\left\{\sum_{i,j}\theta_{i,j}(a^{\dag}_ia_j-b^{\dag}_ib_j)
+U\sum_{i,j}\epsilon_{i,j}I^+_iI^-_j\right\}=\sum_iJ^3_i\nonumber
\end{eqnarray}
where
\begin{eqnarray}
\theta_{i,j}=\delta_{i,j-1}-\delta_{i,j+1},\ \
\epsilon_{ij}=\left\{\begin{array}{cl}1&i<j\\0&i=j\\-1&i>j\end{array}
\right.\label{p3.9}
\end{eqnarray}
By virtue of eqs.~(\ref{p3.3})-(\ref{p3.5}) the validity of eq.~(\ref{p2.21})
for eqs.~(\ref{p3.7})-(\ref{p3.9}) is convinced. One then able to set
\begin{equation}
T^{(1)}_{\pm}=\alpha_{\pm}I_{\pm},\ T^{(1)}_3=2\alpha_+\alpha_-I_3,\
T^{(2)}_{\pm}=\beta_{\pm}J_{\pm},\ T^{(2)}_3=2\alpha_+\beta_-J_3\label{p3.101}
\end{equation}
with
\begin{equation}
\alpha_+\alpha_-=\mu,\ \ \beta_+\alpha_-=\alpha_+\beta_-. \label{p3.111}
\end{equation}
We thus have made a physical realization of eq.~(\ref{p2.62}). Since the
construction of $T^{(n)}_0(\mu)$ has been made the eq.~(\ref{p2.63}) hold,
we should show the validity of eq.~(\ref{p2.66}) for
eqs.~(\ref{p3.6})-(\ref{p3.9}).
Substituting eqs.~(\ref{p3.101}) and (\ref{p3.111}) into eq.~(\ref{p2.66}) and
calculating order by order we find the forms of the formulae of Drinfeld for
$m=n=2$:
\begin{eqnarray*}
[J_{\pm},[J_3,J_{\pm}]]&=&\frac{\mu}{\beta_+\beta_-}
(I_3 J_{\pm}-J_3I_{\pm})I_{\pm},\\
{[J_3,[J_+,J_-]]}&=&\frac{\mu}{\beta_+ \beta_-}(I_+ J_- -J_+ I_-)I_3
\end{eqnarray*}
where
\[ \mu\beta^{-1}_+\beta^{-1}_-=U^2. \]
For $ m=n+1$ and $m=n-1$ one can check that eq.~(\ref{p2.65}) identically
vanishes. It means that eq.~(\ref{p2.65}) is satisfied for $m=3,\ n=2$ and
$m=2,\ n=3$. By direct calculation eq.~(\ref{p2.65}) can be shown to be
true for $m=2,\ n=4$ and $m=n=3$.

What is the physical meaning of such a realization of $\{I_{\lambda}\}$ and
$\{I_{\lambda}\}(\lambda=1,2,3)$? The $\{\vec{I}_i\}$ describe the local spin
``rotation''(up-down) and do not shift the fermion located at lattice.
However the operators $\vec{J}_i$ describe the hopping of fermion from the
$i$-th site to the next ones. More explicitly~\cite{re12} eq.~(\ref{p3.8})
can be written as
\begin{eqnarray}
\lefteqn{J_+=\sum_{i,j}(a^{\dag}_ib_{i+1}-a^{\dag}_{i}b_{i-1})-U\sum_{i,j}
\{a^{\dag}_ib_{i}(a^{\dag}_ja_{j}-b^{\dag}_jb_{j})-a^{\dag}_jb_{j}
(a^{\dag}_ia_{i}-b^{\dag}_ib_{i})\},\nonumber}\\
\lefteqn{J_3=\frac 12 \sum_{i,j}(a^{\dag}_ia_{i+1}-b^{\dag}_{i}b_{i+1}
-a^{\dag}_ia_{i-1}+b^{\dag}_{i}b_{i-1})+U\sum_{i,j}a^{\dag}_ib_{i}
b^{\dag}_ja_{j},\label{p3.10}}\\
\lefteqn{J_-=\sum_{i,j}(b^{\dag}_ia_{i+1}-b^{\dag}_{i}a_{i-1})+U\sum_{i,j}
\{b^{\dag}_ia_{i}(a^{\dag}_ja_{j}-b^{\dag}_jb_{j})-b^{\dag}_ja_{j}
(a^{\dag}_ia_{i}-b^{\dag}_ib_{i})\}\nonumber.}
\end{eqnarray}
The realization eq.~(\ref{p3.10}) possesses a nice property in that
the Hamiltonian of the 1-dim. Hubbard model
\begin{equation}
H=\sum_i(a^{\dag}_ia_{i+1}+a^{\dag}_{i+1}a_{i}+b^{\dag}_{i}b_{i+1}+
b^{\dag}_{i+1}b_{i})-U\sum_i(a^{\dag}_{i}a_{i}-\frac 12)
(b^{\dag}_{i}b_{i}-\frac 12)\label{p3.11}
\end{equation}
commutes with $\{I_{\lambda}\}$ and $\{J_{\lambda}\}$ given by
eqs.~(\ref{p3.6})
and (\ref{p3.10}):
\begin{eqnarray}
[H, I_{\alpha}]&=&0,\label{p3.12}\\
{[H, J_{\alpha}]}&=&0,\ (\alpha=\pm,3)\label{p3.13}
\end{eqnarray}
Therefore
\begin{equation}
[H, Y(sl(2))]=0\label{p3.14}
\end{equation}
i.e., $Y(sl(2))$ serves as a symmetry of 1-d Hubbard model. The
eq.~(\ref{p3.12})
is easy to be proved. For eq.~(\ref{p3.13}) the periodic boundary
condition should be taken account. By substituting eqs.~(\ref{p3.10}) and
(\ref{p3.11}) into the commutator eq.~(\ref{p3.13}) the direct check verifies
the validity of the statement~\cite{re12}.

For a nonlinear model with hopping there exists the interaction between
different sites. Is there a global symmetry besides the ``total'' spin?
The answer is yes. This new symmetry is nothing but the $Y(sl(2))$ in
the example. To form a {\em Yangian} the operators $\{J_{\lambda}\}$
play the central role. They describe the hopping from $i$-th site to
$(i+1)$ or $(i-1)$ plus the operators $\sum_{i\neq j}\vec{S}_i\times
\vec{S}_j$. Obviously $\{J_{\lambda}\}$ cannot be closed, because
the set $\{J_{\lambda}\}$ shifts the fermion one by one site. Only at
the periodic boundary it makes the ``particle'' return to the starting one.
The existence of {\em Yangian} means that we can move a ``particle'' to
any site through only two types of operations:
``local rotation''($\{I_{\lambda}\}$) and ``one-site shift''
($\{J_{\lambda}\}$). The consistence of associativity for the successive
shift operation yields the constraint condition to $\{I_{\lambda}\}$
and $\{J_{\lambda}\}$, as shown by eq.~(\ref{p2.36}). For a physical system
if there is $sl(2)$(or $SU(2)$) symmetry in the absence of interaction, we
may ask whether there exists symmetries that describe the nonlinear-interaction
systems. For the particular classes of models the answer is yes. The
{\em Yangian} is just served as such a symmetry.

\subsection{Example for $Y(sl(n))(n>2)$}

Let us discuss an example for long-ranged interaction model associated with
$sl(n)(n>2)$~\cite{re13}. Introducing the spin operators
\begin{equation}
I_a=\sum^N_{i=1}S^a_i, \ (a=1,\ldots,n)\label{p3.15}
\end{equation}
where
\begin{equation}
[S^a_i, S^b_j]=f_{abc}S^c_i\delta_{ij}\label{p3.16}
\end{equation}
and
\begin{equation}
J_a=\sum_{i\neq j}W_{ij}f_{abc}S^b_iS^c_j,\ (W_{ij}=-W_{ji})\label{p3.17}
\end{equation}
The structure constants $f_{abc}$ obey the Jacobi identity:
\begin{equation}
f_{abc}f_{dec}=f_{adc}f_{bec}-f_{aec}f_{bdc}.\label{p3.18}
\end{equation}
It is easy to verify that eq.~(\ref{p2.41}) is satisfied with
$C_{abc}=f_{abc}$.
As pointed by Drinfeld for $sl(n)(n>2)$ eqs.~(\ref{p2.41}) and
(\ref{p2.42}) imply eq.~(\ref{p2.43}). Therefore we should check the validity
of eq.~(\ref{p2.42}) based on eqs.~(\ref{p3.15}) and (\ref{p3.17}). After
calculation we get
\begin{equation}
[J_a, J_b]=\sum^N_{i\neq j\neq k\neq i}W_{ij}W_{ik}f_{\alpha\beta\mu}
f_{b\gamma\nu}f_{\alpha\mu\nu}T^{\alpha\beta\gamma}_{ijk}\label{p3.19}
\end{equation}
where
\begin{equation}
T^{\alpha\beta\gamma}_{ijk}=S^{\beta}_jS^{\alpha}_iS^{\gamma}_k+S^{\beta}_j
S^{\gamma}_kS^{\alpha}_i+S^{\gamma}_kS^{\alpha}_iS^{\beta}_j
+S^{\alpha}_iS^{\gamma}_kS^{\beta}_j\label{p3.20}
\end{equation}
Hence
\begin{equation}
[J_a,[J_b,J_c]]=f_{bc\lambda}[J_a,J_{\lambda}]=\sum^N_{i\neq j\neq k\neq i}
W_{ij}W_{ik}g^{\alpha\beta\gamma}_{abc}T^{\alpha\beta\gamma}_{ijk}\label{p3.21}
\end{equation}
where
\begin{equation}
g^{\alpha\beta\gamma}_{abc}=f_{\alpha\mu\nu}f_{a\beta\mu}f_{bc\lambda}
f_{\lambda\gamma\nu}\label{p3.22}
\end{equation}
repeated indices are summed. One can prove that
\begin{equation}
(g^{\alpha\beta\gamma}_{abc}+g^{\alpha\beta\gamma}_{bca}+g^{\alpha\beta\gamma}
_{cab})+(g^{\alpha\gamma\beta}_{abc}+g^{\alpha\gamma\beta}_{bca}+
g^{\alpha\gamma\beta}_{cab})=\sum t^{(\alpha\beta\gamma)}_{abc}\label{p3.23}
\end{equation}
where $(\alpha\beta\gamma)$ denotes the summation over the permutation to
$\alpha,\ \beta$ and $\gamma$
\begin{equation}
t^{\alpha\beta\gamma}_{abc}=f_{a\alpha\lambda}f_{b\beta\mu}f_{c\gamma\nu}
f_{\lambda\mu\nu}=4!\ a_{abc\alpha\beta\gamma} \label{p3.24}
\end{equation}
that has appeared in eq.~(\ref{p2.37}).

With the notations we have
\begin{equation}
\sum^N_{i\neq j\neq k\neq i}W_{ij}W_{ik}
(g^{\alpha\beta\gamma}_{abc}+g^{\alpha\beta\gamma}_{bca}+
g^{\alpha\beta\gamma}_{cab})T^{\alpha\beta\gamma}_{ijk}
=\sum^N_{i\neq j\neq k\neq i}W_{ij}W_{ik}t^{(\alpha\beta\gamma)}_{abc}
T^{\alpha\beta\gamma}_{ijk}\label{p3.25}
\end{equation}
With the help of the above relations on the basis of the Jacobi identities
\begin{equation}
[J_a,[J_b,I_c]]+[I_a,[J_b,J_c]]=[J_a,[J_b,I_c]]+[J_c,[I_a,J_b]]+
[J_b,[J_c,I_a]]. \label{p3.26}
\end{equation}
Only the case where $a\neq b\neq c \neq a$ should be taken into account.
These components of eq.~(\ref{p3.26}) are equal to
\begin{eqnarray}
2&\sum^N_{i\neq j\neq k}&(\ W_{ij}W_{ik}t^{\alpha\beta\gamma}_{abc}S^{\alpha}_i
S^{\beta}_jS^{\gamma}_k+W_{ik}W_{ij}t^{\alpha\beta\gamma}_{abc}S^{\alpha}_i
S^{\gamma}_kS^{\beta}_j\nonumber\\
&&+W_{ki}W_{kj}t^{\alpha\beta\gamma}_{abc}S^{\gamma}_k
S^{\alpha}_iS^{\beta}_j+W_{ji}W_{jk}t^{\alpha\beta\gamma}_{abc}S^{\beta}_j
S^{\alpha}_iS^{\gamma}_k\nonumber\\
&&+W_{jk}W_{ji}t^{\alpha\beta\gamma}_{abc}S^{\beta}_j
S^{\gamma}_kS^{\alpha}_i+W_{kj}W_{ki}t^{\alpha\beta\gamma}_{abc}S^{\gamma}_k
S^{\beta}_jS^{\alpha}_i)\nonumber\\
&=&\frac 23\sum^N_{i\neq j\neq k\neq i}(W_{ij}W_{ik}+W_{jk}W_{ji}+W_{ki}W_{kj})
t^{\alpha\beta\gamma}_{abc}\{S^{\alpha}_i,S^{\beta}_j,S^{\gamma}_k\}\label{3.27}
\end{eqnarray}
Defining
\begin{equation}
\Delta_{ijk}=W_{ij}W_{ik}+W_{jk}W_{ji}+W_{ki}W_{kj},\
(W_{ij}=-W_{ji})\label{p3.28}
\end{equation}
and noting that there is no summation over the repeated lattice indices, when
\begin{equation}
\Delta_{ijk}=h^2,\ \ (\mbox{independent of}\ i,j,k)\label{p3.29}
\end{equation}
the lhs of eq.~(\ref{p3.26})(with $a\neq b\neq c \neq a$) is equal to
\begin{equation}
h^2a_{abcdef}\{I^d,I^e,I^f\}\label{p3.30}
\end{equation}
which is exactly  the relation eq.~(\ref{p2.42}). Thus let eq.~(\ref{p3.17})
satisfy $Y(sl(2))$ the eq.~(\ref{p3.28}) should be satisfied, i.e.,
\begin{equation}
\Delta_{ijk}=W_{ij}W_{ik}+W_{jk}W_{ji}+W_{ki}W_{kj}=h^2,\ (\forall\ i\neq j\neq
k
\neq i)\label{p3.31}
\end{equation}
which is the same as eq.~(\ref{pw}). Besides the solution (\ref{pw1}),
the general solution~\cite{re14}
\begin{equation}
W_{jk}=(i h)\frac{Z_j+Z_k}{Z_j-Z_k}\label{p3.32}
\end{equation}
where $\{Z_j\}$ is a discrete set of complex number. Thus
\begin{equation}
J_a=U\sum^N_{i=1}S^a_i+ih\sum^N_{j\neq k}\frac{Z_j+Z_k}{Z_j-Z_k}
f_{abc}S^b_jS^c_k,\ (a,b,c=1,2,\ldots,n)\label{p3.33}
\end{equation}
together with eq.~(\ref{p3.15}) generate the $Y(sl(n))$ for a free parameter
$U$.
Here we have used the important property of {\em Yangian}~\cite{re10,re14}:

Suppose $\{I_a\}$ and $\{J_a\}$ generate {\em Yangian} then the $\{I_a\}$ and
$\{\tilde{J}_a\}$ also generate a {\em Yangian}, where
\begin{equation}
\tilde{J}_a=UI_a+J_a,\label{p3.34}
\end{equation}
$U$ being an arbitrary parameter. This statement can be verified easily because
$\tilde{J}_a$ behaves completely in the same manner as $J_a$.

The simplest form of $W_{jk}$ is $Z_k=e^{i\frac{k\pi}N}$($k$ integer),
then
\begin{equation}
W_{jk}=h\coth\frac{(j-k)\pi}N.\label{p3.35}
\end{equation}
The associated Hamiltonian systems were found in Refs.~\cite{re15}.
Now we can deduce the Hamiltonian from our point of view.

Noting that
\begin{equation}
C_2=T^{(2)}_0-\frac 12(2I^2_3+I_+I_-+I_-I_+)+\frac 12[T^{(1)}+\frac 12
(T^{(1)}_0)^2]\end{equation}
Assume
\begin{equation}
T^{(2)}_0=\sum_{i\neq j}f_{ij}(P_{ij}-1)
\end{equation}
where $P_{ij}$ stand for the permutation operators
\begin{equation}
P_{ij}=2 S^3_iS^3_j+S^+_iS^-_j+S^-_iS^+_j-\frac 12
\end{equation}
then from
\begin{eqnarray*}
[T^{(2)}_0,T^{(2)}_{\pm}]&=&\pm[T^{(1)}_3T^{(2)}_{\pm}-T^{(2)}_3T^{(1)}_{\pm}]\\
{[T^{(2)}_0,T^{(2)}_3]}&=&2[T^{(1)}_+T^{(2)}_--T^{(2)}_+T^{(1)}_-]
\end{eqnarray*}
it follows
\[f_{ij}=\frac 12 W^2_{ij}\]
we obtain
\begin{equation}
C_2=-2H_2+N(N-1)+\frac 12 [T^{(1)}_0+\frac 12(T^{(1)}_0)^2]
\end{equation}
where
\begin{equation}
H_2=-\frac 14\sum_{i\neq j}(W^2_{ij}-1)(P_{ij}-1)
\end{equation}
and $T^{(1)}_0,\ C_2,\  \{I_{\alpha}\}$ and $\{J_{\alpha}\}(\alpha=\pm,\ 3)$
are all commutative. Therefore we have the Hamiltonian
\begin{equation}
H_2=\sum_{i\neq j}\frac{Z_iZ_j}{Z_{ij}Z_{ji}}(P_{ij}-1),\ (Z_{ij}=Z_i-Z-j)
\end{equation}
that was obtained in Ref.~\cite{re15}

\section{Long-Range Interaction Models and {\em Yangian}}
\setcounter{equation}{0}

\subsection{Polychronakos' Approach}

Besides the nearest neighborhood models(such as Heisenberg chain, Hubbard
model, $\ldots$), recently, a number of one-dimensional long-range interaction
models have been studied~\cite{re15}.
The typical one is Calogero-Sutherland model~\cite{re16},
then it is subsequently extended to the models with internal spin
degrees of freedom~\cite{re17}. Among them an interesting approach
was proposed by Bernard, Gaudin, Haldane and Pasquier (BGHP)\cite{re17}
who made this type of models related to the RTT relation associated with
Yang-Baxter equation(YBE).
The BGHP approach provides a method to deal with long-range interaction
models: for a given rational solution of YBE,
for example, $R(u)=u+P$ which is given in eq.~(\ref{p1.18}),
RTT relation gives rise to the {\em Yangian} symmetry.
With a particular realization
of the {\em Yangian}, in general, we can generate corresponding Hamiltonian
of the considered systems.

On the other hand Polychronakos had formulated the integrability in terms
of the ``coupled'' momentum operators~\cite{re18}:
\begin{equation}
\pi_i=p_i+i\sum_{j\neq i}V_{ij}K_{ij} \label{pi}
\end{equation}
where $p_i=-i\frac{\partial}{\partial x_i}\ (\hbar=1)$,
$V_{ij}=V(x_i-x_j)$ a potential to be determined and $K_{ij}$ the
particle permutation operators. The requirements of the Hermiticity of
$\pi_i$, the absence of linear terms in $p_i$ and that only the
two-body potentials in the Hamiltonian lead to
\begin{eqnarray}
\lefteqn{V(x) = -V(-x)\ ,} \nonumber\\
\lefteqn{H_0 \equiv \frac 1 2 \sum_i \pi^2_i=\sum_ip^2_i+\frac 1 2 \sum_{i\neq
j}
\left[ \frac{\partial}{\partial x_i} V_{ij}K_{ij}
+V^2_{ij}\right] - \frac 1 6 \sum_{i\neq j \neq k \neq i}V_{ijk}K_{ijk}}
\end{eqnarray}
where
\begin{eqnarray}
V_{ijk} &=& V_{ij}V_{jk}+V_{jk}V_{ki}+V_{ki}V_{ij} = W_{ij}+W_{jk}+W_{ki}\ ,\\
K_{ijk} &=& K_{ij}K_{jk}\nonumber
\end{eqnarray}
with $W_{ij}=W(x_i-x_j)$ being a symmetric function.
The commutation relation between $\pi_i$ and $\pi_j$ is found to be
\begin{equation}
[\pi_i,\pi_j]=\sum_{k\neq i,j}V_{ijk}(K_{ijk}-K_{jik})\ .
\end{equation}
This approach can be applied to many integrable systems, including the
CS model, $\delta$-interaction model and so on.

\subsection{Sutherland-R\"omer and Yan Models}

Let us first discuss the extended forms of $V_{ij}$ in eq.~(\ref{pi}) that
are different from those given by Ref.~\cite{re18}. Setting
\begin{equation}
V_{ij}=P^+_{ij}a_{ij}+P^-_{ij}b_{ij}\label{vij}
\end{equation}
where $P^{\pm}_{ij}$ are given as
\begin{equation}
P^{\pm}_{ij}=\frac{1\pm \sigma_i\sigma_j}2\ (\sigma^2_i=1)\label{pij}
\end{equation}
$\sigma_i$ is quantum operators obeying
\begin{eqnarray}
\sigma_iK_{ij} &=& K_{ij}\sigma_j\ ,\nonumber\\
\sigma_iK_{mn} &=& K_{mn}\sigma_i\ \ (i\neq m ,n)\ ,\nonumber
\end{eqnarray}
then by substituting eq.~(\ref{vij}) into eq.~(\ref{pi}) and doing
the parallel discussion in Ref.~\cite{re18}, we find
\begin{equation}
V_{ijk}=P^+_{ijk}A_{ijk}+P^-_{ijk}A_{ijk}+P^-_{kij}A_{kij}+P^-_{jki}A_{jki}
\end{equation}
where
\begin{eqnarray}
P^{\pm}_{ijk} &=& P^{\pm}_{ij}P^{\pm}_{ik}\ ,\nonumber\\
A_{ijk} &=& a_{ij}a_{jk}+a_{jk}a_{ki}+a_{ki}a_{ij}\ ,\nonumber\\
B_{ijk} &=& a_{ij}b_{jk}+b_{jk}a_{ki}+b_{ki}a_{ij}\ .\nonumber
\end{eqnarray}
Noting that
$P^+_{ijk}=P^+_{ikj}=\cdots=P^+_{kji}$, but $P^-_{ijk}=P^-_{jik}$ only.

The sufficient condition of the quantum integrability of
eq.~(\ref{pi}) is
\begin{equation}
V_{ijk}=\mbox{constant} \ \ (\mbox{or \ zero}) \label{suf} \ .
\end{equation}

Now let us look for new solution of eq.~(\ref{suf})

\noindent{\bf (1)} When $A_{ijk}\neq 0$, $B_{ijk}\neq 0$,
a sufficient solution can be checked:
\begin{eqnarray}
a(x)=l\coth(ax) \ \ &&(\mbox{or} \ a(x)=l\cot(ax))\ ,\nonumber\\
b(x)=l\tanh (ax) \ \ &&(\mbox{or} \ b(x)=l\tan(ax))
\end{eqnarray}
where $x \equiv x_{ij}=x_i-x_j$, $a,\ l$ are constants and
\begin{equation}
V_{ijk}=-l^2(P^+_{ijk}+P^-_{ijk}+P^-_{kij}+P^-_{jki})=-l^2\ .\label{vv}
\end{equation}
Define
\begin{equation}
H=\frac 12 \sum_i \pi^2_i-\frac {l^2}{6} \sum_{i\neq j \neq k \neq i}K_{ijk}\ ,
\end{equation}
then eq.~(\ref{vv}) leads to
\begin{equation}
H=\frac 12 \sum_i p^2_i+\sum_{i<j}l(l-aK_{ij})\left[\frac{P^+_{ij}}
{\sinh^2(ax_{ij})}-\frac{P^-_{ij}}{\cosh^2(ax_{ij})}\right]\ . \label{hrs}
\end{equation}
When $K_{ij}=\pm 1$, eq.~(\ref{hrs}) was firstly given in Ref.~\cite{r19}
and studied by Sutherland and R\"omer~\cite{re19} in detail.

Define
\begin{equation}
\bar {\pi}_i=\pi_i+il\sum_{i\neq j}K_{ij}\ ,
\end{equation}
then
\begin{eqnarray}
[\bar {\pi}_i, \bar {\pi}_j] & = & 2il(\bar {\pi}_i- \bar {\pi}_j)K_{ij}\\
{[ H, \pi_i]} &=& [H, \bar {\pi}_i]  =  0\ .
\end{eqnarray}
The conserved quantities are given by
\begin{equation}
I_n=\sum_i\bar{\pi}^n_i
\end{equation}
which leads to
\begin{eqnarray}
[I_n, I_m] &=& 0\ ,\label{ii}\\
{[H, I_n]} &=& 0\ ,
\end{eqnarray}
i.e. the model is quantum integrable in the sense of Liouville.

\noindent{\bf (2)} When $B_{ijk}=0$, we consider two cases

\noindent{\bf (a)}\hspace*{0.5cm} $A_{ijk}=0$
\begin{eqnarray}
a(x) &=& \frac l x\ ,\ \  V_{ijk}=0\ ,\nonumber\\
H &=& \frac 12 \sum_ip^2_i + \frac 12 \sum_{i\neq j}
       \frac{l(l-K_{ij})}{(x_i-x_j)^2}P^+_{ij}
\end{eqnarray}
that is well known as Calogero model
when $P^+_{ij}$ takes the value $1$.

\noindent{\bf (b)}\hspace*{0.4cm}$A_{ijk}=\beta^2 \neq 0$
\begin{equation}
[\pi_i, \pi_j]=\beta \sum_{k\neq i,i}P^+_{ijk}(K_{ijk}-K_{jik}) \label{pic}\ .
\end{equation}
Define
\begin{equation}
\bar {\pi}_i=\pi_i +\beta \sum_{i\neq j}P^+_{ij}K_{ij}\label{bpi}\ ,
\end{equation}
it is easy to prove that
\begin{equation}
[\bar {\pi}_i, P^+_{jk}]=0, \ \ \forall i\  \mbox{and}\ j\neq k
\end{equation}
and
\begin{eqnarray}
[\bar {\pi}_i, \bar {\pi}_j] &=& 2\beta P^+_{ij}
(\bar {\pi}_i-\bar {\pi}_j)K_{ij}\label{pi2}\ ,\\
{[ \bar {\pi}^n_i, \bar {\pi}_j]} &=& 2\beta P^+_{ij}
(\bar {\pi}^n_i-\bar {\pi}^n_j)K_{ij}\ ,
\end{eqnarray}
so that eq.~(\ref{ii}) is also satisfied. Let
\begin{equation}
H=\frac 12 \sum_i \pi^2_i+\frac {\beta^2}6 \sum_{i\neq j\neq k\neq i}
P^+_{ijk}K_{ijk}\ .
\end{equation}
With the help of eq.~(\ref{pic}), one can prove
\begin{equation}
[H, \pi_i]=[H, \bar {\pi}_i]=[H, I_n]=0\ .
\end{equation}

For the case (b) we have two sufficient solutions of $V_{ijk}$:

\noindent (i)\hspace*{0.6cm}$a(x)=il\cot(ax) \ \ (\mbox{or} a(x)=l\coth(ax)$,
\begin{eqnarray}
\lefteqn{V_{ijk}=-l^2P^+_{ijk}\ ,\nonumber}\\
\lefteqn{H=\frac 12 \sum_i  p^2_i +\frac 1 2 \sum_{i\neq j}
\frac{l(l-aK_{ij})}{\sin^2a(x_i-x_j)}P^+_{ij}\ . \label{sc}}
\end{eqnarray}
Eq.~(\ref{sc}) is the generalization of the spin chain model
considered by BGHP~\cite{re17}.

\noindent (ii)\hspace*{0.45cm}$a(x)=l\mbox{sgn}(x)$,
\begin{equation}
H=\frac 12\sum_i p^2_i+\frac 12 \sum_{i\neq j}l(l-K_{ij})
\delta(x_i-x_j)P^+_{ij}\ .\label{hd}
\end{equation}
with the condition that $K_{ij}=\pm 1$, eq.~(\ref{hd}) was first pointed out
by Yan~\cite{re20} through Bethe Ansatz, he also found the Y-operator
defined by Yang~\cite{re6} for eq.~(\ref{hd})
\begin{equation}
Y^{\alpha \beta}_{ij}=\frac{1}{ik_{ij}(ik_{ij}-2c)}[ik_{ij}-c(1-
\sigma_i\sigma_j)][-ik_{ij}P^{\alpha \beta}+c(1+\sigma_i \sigma_j)]
\end{equation}
where $P$ is the permutation and $Y$ satisfies
\begin{equation}
Y^{\alpha \beta}_{jk}Y^{\beta \gamma}_{ik}Y^{\alpha \beta}_{ij}=
Y^{\beta \gamma}_{ij}Y^{\alpha \beta}_{ik}Y^{\beta \gamma}_{jk}
\end{equation}
and $c=l(l\pm 1)/2$ for $K_{ij}=\pm 1$. Noting that there is only
$P^+_{ij}$ in the Hamiltonian eq.~(\ref{hd}) for the quantum
integrability.

Now we have re-interpreted the models eq.~(\ref{hrs})
and eq.~(\ref{hd}) from the point of view of the
formulation eq.~(\ref{pi}). Next we shall set up the
{\em Yangian} description of models eq.~(\ref{hrs})
and eq.~(\ref{hd}) through RTT relation.

\subsection{RTT Relation and Long-Range Interaction Models}

Let us apply the BGHP approach~\cite{re17} to the S-R model and Yan model.
The solution of Yang-Baxter equation, $R$-matrix, takes the
simplest form~(\ref{p1.18}) as
\begin{equation}
R(u)=u+\lambda P_{00^{\prime}}\label{r}
\end{equation}
and the RTT relation reads
\begin{equation}
R_{00^{\prime}}(u-v)T^0(u)T^{0^{\prime}}(v)=
T^{0^{\prime}}(v)T^0(u)R_{00^{\prime}}(u-v)\label{rtt}
\end{equation}
where $T^0(u)=T(u)\otimes 1$, $T^{0^{\prime}}=1\otimes T(u)$ and
$P_{00^{\prime}}$ is the permutation operator exchanging the two auxiliary
spaces $0$ and $0^{\prime}$. Make the expansion
\begin{eqnarray}
T^0(u) &=& I+\sum^p_{a,b=1}X^0_{ba}\sum^{\infty}_{n=0}
           \frac{\lambda T^{ab}_n}{u^{n+1}}\ ,\label{t}\\
P_{00^{\prime}} &=& \sum^p_{a,b=1}X^0_{ba}X^{0^{\prime}}_{ab}\ .\label{p}
\end{eqnarray}
Notice that $T^{ab}_n$ is the same as $T^{(n+1)}_{ba}$ in the \S1 and \S2.
It is well known that $\{T^{ab}_n\}$ generate the {\em Yangian}~\cite{re10}.
Substituting eqs.~(\ref{r}),~(\ref{t}) and (\ref{p}) into eq.~(\ref{rtt})
one finds
\begin{equation}
\hspace*{-0.5cm}\sum_{a,b}\sum_{c,d}X^0_{ba}X^{0^{\prime}}_{dc}\sum^{\infty}
 _{n=0} \left\{ u^{-n-1}f^n_1-v^{-n-1}f^n_2 + \sum^{\infty}_{m=0}u^{-n-1}
 v^{-m-1}f^{n,m}_3 \right\} =0
\end{equation}
where
\begin{eqnarray}
f^n_1 &=& \delta_{bc}T^{ad}_n-\delta_{ad}T^{cb}_n-[T^{ab}_n,T^{cd}_0]\ ,
    \nonumber\\
f^n_2 &=& \delta_{bc}T^{ad}_n-\delta_{ad}T^{cb}_n-[T^{ab}_0,T^{cd}_n]\ ,
     \nonumber\\
f^{n,m}_3 &=& \lambda(T^{ad}_nT^{cb}_m-T^{ad}_mT^{cb}_n)+
[T^{ab}_{n+1},T^{cd}_m]-[T^{ab}_{n},T^{cd}_{m+1}]\ .\nonumber
\end{eqnarray}
For any auxiliary space $\{X_{ab}\}$ we require $f^n_1=f^n_2=f^{n,m}_3=0$.
Obviously, $f^n_1=0$ is equivalent to $f^n_2=0$. So we need only to take
\begin{equation}
f^n_1=f^{n,m}_3=0\label{f}
\end{equation}
into account.

First from $f^{n,0}_3=0$ it follows
\begin{equation}
\delta_{bc}T^{ad}_{n+1}-\delta_{ad}T^{cb}_{n+1}=\lambda (T^{ad}_0T^{cb}_n
-T^{ad}_nT^{cb}_0)+[T^{ab}_n, T^{cd}_1]
\end{equation}
which can be recast to
\begin{eqnarray}
\lefteqn{T^{ad}_{n+1} = \lambda(T^{ad}_0T^{cc}_n-T^{ad}_nT^{cc}_0)+[T^{ac}_n,
   T^{cd}_1]\ (a\neq d)\ , \label{t4}}\\
\lefteqn{T^{aa}_{n+1}-T^{cc}_{n+1} = \lambda(T^{aa}_0T^{cc}_n-T^{aa}_nT^{cc}_0)
   +[T^{ac}_n, T^{ca}_1]\ , \label{t2}}
\end{eqnarray}
where no summation for the repeating indices is taken. Eqs.~(\ref{t4}) and
(\ref{t2}) imply that $T^{ab}_n$ can be determined by iteration for
given $T^{ab}_0$ and $T^{ab}_1$.

Now let us set
\begin{eqnarray}
T^{ab}_0 &=& \sum^N_{i=1}I^{ab}_i\ , \label{ti} \\
T^{ab}_1 &=& \sum^N_{i=1}I^{ab}_iD_i \label{d}
\end{eqnarray}
and
\begin{equation}
[I^{ab}_i, I^{cd}_j]= \delta_{ij}(\delta_{bc}I^{ad}_i-\delta_{ad}I^{cb}_i)
\label{ij}
\end{equation}
where $D_i$ are operators to be determined.  Substituting
eqs.~(\ref{ti})--(\ref{ij}) into $f^1_1$ we obtain
\begin{equation}
\sum_i\sum_j I^{ab}_i[D^i, I^{cd}_j]=0\ .
\end{equation}
Further we assume
\begin{equation}
\sum_i I^{ab}_i[D_i, I^{cd}_j]=0, \mbox{for\ any}\ j \label{di}
\end{equation}
with which the $T^{ab}_2$ should satisfy
\begin{eqnarray}
\delta_{bc}T^{ad}_2-\delta_{ad}T^{cb}_2 &=& \sum_{i\neq j}I^{ab}_iI^{cd}_j
    \left\{ \lambda \sum_{k,l}I^{kl}_iI^{lk}_j (D_j-D_i)+[D_i,D_j] \right\}
    \nonumber\\
&&+\sum_i(\delta_{bc}I^{ad}_iD^2_i-\delta_{ad}I^{cb}_iD^2_i)\label{t3}\ .
\end{eqnarray}
A sufficient solution of eq.~(\ref{t3}) is
\begin{equation}
T^{ab}_2=\sum_i I^{ab}_iD^2_i
\end{equation}
with
\begin{equation}
[D_i, D_j]=\lambda \sum_{a,b}I^{ab}_j I^{ba}_i(D_i-D_j)\label{did}\ .
\end{equation}
Thus eq.~(\ref{d}) generates long-range interaction through the
eq.~(\ref{di}) and (\ref{did}).
However so far there is not simple relationship between $D_i$ and $I^{ab}_j$
which should satisfy eq.~(\ref{di}). It is very difficult to determine
the general relationship. Fortunately, BGHP~\cite{re17} have set up the link
with the help of projection. Let the permutation groups $\Sigma_1$,
$\Sigma_2$ and $\Sigma_3$ be generated by $K_{ij}$, $P_{ij}$ and
the product $P_{ij}K_{ij}$ respectively, where $K_{ij}$ exchange
the positions of particles and $P_{ij}$ exchange the spins at position
$i$ and $j$. The projection $\rho$ was defined as
\begin{equation}
\rho (ab)=a \ \ \mbox{for}\ \forall a\in \Sigma_2, b\in \Sigma_1\ ,
\end{equation}
i.e. the wave function considered is symmetric. Let $I^{ab} _i$ be the
fundamental representations, then
\begin{equation}
P_{ij}=\sum_{a,b}I^{ab}_i I^{ba}_j\ .
\end{equation}
Suppose that there exists~\cite{re17}
\begin{equation}
D_i=\rho ({\hat D}_i), \ \ D_i \in \Sigma_2,\ {\hat D}_i\in \Sigma_1
\label{eee}
\end{equation}
and the ${\hat D}_i$ is particle-like operators, i.e.
\begin{equation}
K_{ij}{\hat D}_i={\hat D}_jK_{ij}, \ \ K_{ij}{\hat D}_l={\hat D}_lK_{ij}
\ \  (l\neq i,j)\ .
\end{equation}
Define
\begin{equation}
T^{ab}_m=\sum_i I^{ab}_i \rho ({\hat D}^m_i) \ \ (m\geq 0)\ ,\label{tt}
\end{equation}
then
\begin{equation}
\hspace*{-1.1cm}(\mbox{a})\hspace*{0.7cm}
[{\hat D}_j, {\hat D}_i] = \lambda \rho^{-1}(P_{ij}(D_j-D_i))
        = \lambda ({\hat D}_j-{\hat D}_i)K_{ij}\ . \label{dd}
\end{equation}
(b) $T^{ab}_m$ satisfy eq.~(\ref{f}), i.e., RTT relation eq.~(\ref{rtt}).
Actually $f^n_1=0$ is easy to be checked. By using
\[
[{\hat D}^n_i, {\hat D}^m_j] = \sum^{n-1}_{k=0}{\hat D}^k_i[{\hat D}_i,
       {\hat D}^m_j]{\hat D}^{n-k-1}_j = \lambda \sum^{n-1}_{k=0}
       {\hat D}^k_i({\hat D}^m_i-{\hat D}^m_j){\hat D}^{n-k-1}_jK_{ij}\ ,
\]
we have $f^{n,m}_3=0$.

The projection procedure is very important for it enables us to prove that
eq.~(\ref{f}) is satisfied by virtue of eq.~(\ref{eee}).

With the expansion eqs.~(\ref{t}) and the projected long-range
expansion eq.~(\ref{tt}), the hamiltonian associated to $T(u)$ is
obtained by the expansion of the deformed determinant~\cite{re17}:
\begin{equation}
\mbox{det}_qT(u)=\sum_{\sigma} \epsilon(\sigma)T_{1\sigma_{1}}(u-(p-1)\lambda)
T_{2\sigma_{2}}(u-(p-2)\lambda)\cdots T_{p\sigma_{p}}(u)\ .
\end{equation}
A calculation gives
\begin{eqnarray}
\mbox{det}_qT(u) &=& 1+\frac {\lambda}uM+\frac{\lambda}{u^2}\left[
  \rho(\sum_i{\hat D}_i-\frac{\lambda}2 \sum_{j\neq i}K_{ij})+
  \frac{\lambda}2 M(M-1)\right]\nonumber\\
&& +\frac{\lambda}{u^3}\rho\left\{ (\sum_i{\hat D}_i-\frac{\lambda}2\sum_
  {j\neq i}K_{ij})^2+\frac{\lambda^2}{12}\sum_{i\neq j\neq k\neq i}K_{ij}K_{jk}
  \right. \nonumber\\
&& +\lambda (M-1)\sum_i ({\hat D}_i-\frac{\lambda}2\sum_{j\neq i}K_{ij})
  \nonumber\\
&& +\left.\frac{\lambda^2}6M(M-1)(M-2)+\frac{\lambda^2}4M(M-1)\right\}
   +\cdots\ .\label{det}
\end{eqnarray}
One takes the Hamiltonian as
\begin{equation}
H=\frac 12 \rho \left\{(\sum_i{\hat D}_i-\frac{\lambda}2 \sum_{i\neq j}K_
  {ij})^2+\frac{\lambda^2}{12}\sum_{i\neq j\neq k\neq i}K_{ij}K_{jk}\right\}\ .
\end{equation}
Therefore we define the Hamiltonian which have the {\em Yangian}
symmetry given by eqs.(\ref{tt}), (\ref{ij}) and (\ref{did}).
In comparison to the known models we list the expressions for ${\hat D}_i$
satisfying eq.~(\ref{dd})~\cite{re18}

\noindent{\bf (1)} \hspace*{0.45cm} $ {\hat D}_i = p_i+\frac{\lambda}2
 \sum_{i\neq j}[\mbox{sgn}(x_i-x_j)+1]K_{ij}, \ \ \lambda=2il$,
\begin{equation}
H= \frac 12 \sum_ip^2_i +\frac 12 \sum_{i\neq j}l(l-P_{ij})
      \delta(x_i-x_j)\ .\label{hh1}
\end{equation}

\noindent{\bf (2)} \hspace*{0.45cm} $ {\hat D}_i = p_i+\sum_{i\neq j}l
 [i\cot a(x_i-x_j)+1]K_{ij},\  \lambda=2l$,
\begin{equation}
H = \frac 12 \sum_ip^2_i +\frac 12 \sum_{i\neq j}\frac{l(l-aP_{ij})}
{\sin^2a(x_i-x_j)}\ .\label{hh2}
\end{equation}

\noindent{\bf (3)}\hspace*{0.45cm}${\hat D}_i = p_i+il\sum_{i\neq j}
[\coth a(x_i-x_j)P^+_{ij}+\tanh a(x_i-x_j)P^-_{ij}+1]K_{ij}, \ \lambda=2il\ , $
\begin{equation}
H = \frac 12 \sum_ip^2_i +\frac 12 \sum_{i\neq j}l(l-aP_{ij})
 \left(\frac{P^+_{ij}}{\sinh^2a(x_i-x_j)}-\frac{P^-_{ij}}{\cosh^2a(x_i-x_j)}
 \right).\label{h3}
\end{equation}
Eqs.~(\ref{hh1}) and (\ref{hh2}) were given in Ref.~\cite{re18},
eq.~(\ref{hh2}) was studied in Ref.~\cite{re17}. Eq~(\ref{h3})
is the generalization of S-R model.

An alternative description of transfer matrix was given by BGHP~\cite{re17}.
Define
\begin{equation}
{\bar D}_i={\hat D}_i-\lambda \sum_{i<j}K_{ij}\ ,
\end{equation}
then
\begin{eqnarray}
\lefteqn{[{\bar D}_i, {\bar D}_j] =0\ , \label{d1}}\\
\lefteqn{[ K_{ij},{\bar D}_k ] =0\ \ (k\neq i,j)\ ,\label{d2}}\\
\lefteqn{K_{ij}{\bar D}_i-{\bar D}_jK_{ij} = \lambda\ .}
\end{eqnarray}
It was proved that
\begin{equation}
{\bar T}_i(u)=1+\lambda \frac{P_{0i}}{u-{\bar D}_i},\ {\bar T}(u)
=\prod_i{\bar T}_i(u)\ \mbox{and}\ \rho({\bar T}(u))\label{t1}
\end{equation}
all satisfy the RTT relation.

The deformed determinant of ${\bar T}(u)$ was defined by
\begin{equation}
\mbox{det}_q{\bar T}(u)=\frac{\Delta_M(u+\lambda)}{\Delta_m(u)}\ ,\ \
\Delta_M(u)=\prod^M_{i=1}(u-{\bar D}_i)\ .
\end{equation}
It was proved that
\begin{equation}
\rho(\mbox{det}_q{\bar T}(u))=\mbox{det}_q(T(u))\ .
\end{equation}

To contain the model eq.(\ref{hd}), we define ${\bar D}_i$
related to the ${\bar \pi}_i$ given by eq.~(\ref{bpi}) as
\begin{equation}
{\bar D}_i={\bar \pi}_i-\beta\sum_{j<i}P^+_{ij}K_{ij}
\end{equation}
which satisfies eqs.(\ref{d1}), (\ref{d2}) and (\ref{t1}) etc. So
we can put the models eqs.~(\ref{hrs}) and (\ref{hd}) into
Yang-Baxter system.

In conclusion of this section we have shown the consistence between
{\em Yangian} symmetry and the integrability of Polychronakos for
long-range interaction models \cite{re18} and given the interpretation of
S-R model~\cite{r19,re19}
and Yan model~\cite{re20} from the point of view of YB system.

\section{Truncated Transfer Matrix $T(u)$ and G-C Top}
\setcounter{equation}{0}

For the $R$-matrix is given by eq.~(\ref{p1.18}) the case (A) in the section
II yields {\em Yangian}. Now let's look for what kind of algebra appears for
the
case (B), i.e., $T^{(0)}=\left[\begin{array}{cc}1&0\\0&0\end{array}\right]$.
In this section the transfer matrix for the case (B) is denoted by $t(u)$,
the RTT relation is calculated to give the following relations($a,b=1,2$).
\begin{equation}
[t^{(m)}_{ab}, t^{(k)}_{ab}]=0,\ \ [t^{(m)}_{ab}, t^{(k)}_{cd}]=
[t^{(k)}_{ab}, t^{(m)}_{cd}] ,\ (m,k\geq 1)\label{p5.1}
\end{equation}
that are the same as those for the case (A), the other relations are given by
\begin{eqnarray}
\lefteqn{\left\{\begin{array}{l}   [t^{(1)}_{11},t^{(k)}_{22}]=[t^{(1)}_{12},
t^{(k)}_{22}]=[t^{(1)}_{21},t^{(k)}_{22}]=0,\\ t^{(k)}_{12}=[t^{(k)}_{12},
t^{(1)}_{11}],\ t^{(k)}_{21}=-[t^{(k)}_{21}, t^{(1)}_{11}],\ t^{(k)}_{22}=
[t^{(k)}_{12},t^{(1)}_{21}],\ (k\geq 1)\end{array}\right.\label{p5.2}}\\
\lefteqn{\left\{\begin{array}{l}  [t^{(2)}_{22},t^{(k)}_{11}]+t^{(k)}_{21}
t^{(1)}_{12}-t^{(1)}_{21}t^{(k)}_{12}=0,\\{[t^{(2)}_{12},t^{(k)}_{22}]}+
t^{(k)}_{12}t^{(1)}_{22}-t^{(1)}_{12}t^{(k)}_{22}=0,\ (k\geq 1)\\
{[t^{(2)}_{21},t^{(k)}_{22}]}+t^{(k)}_{22}t^{(1)}_{21}-t^{(1)}_{22}
t^{(k)}_{21}=0 \end{array}\right.} \label{p5.3}
\end{eqnarray}
and the iteration relations
\begin{eqnarray}
t^{(k+1)}_{12}&=&[t^{(2)}_{12},t^{(k)}_{11}]+t^{(k)}_{11}t^{(1)}_{12}
-t^{(1)}_{11}t^{(k)}_{12},\nonumber \\
t^{(k+1)}_{21}&=&[t^{(k)}_{11},t^{(2)}_{21}]+t^{(k)}_{11}t^{(1)}_{21}
-t^{(1)}_{11}t^{(k)}_{21},\ (k\geq 2) \label{p5.4}\\
t^{(k+1)}_{22}&=&[t^{(2)}_{12},t^{(k)}_{21}]+t^{(k)}_{11}t^{(1)}_{21}
-t^{(1)}_{11}t^{(k)}_{22}.\nonumber
\end{eqnarray}
Not all the above relations are independent. By substituting eq.~(\ref{p5.4})
into eq.~(\ref{p5.2}) and taking eq.~(\ref{p5.1}) into account it can be shown
that only $k=1,2$ in eq.~(\ref{p5.2}) is independent. Thus eq.~(\ref{p5.2}) is
simplified to\\

\hspace*{0.5cm}
$\begin{array}{ll}[t^{(k)}_{12}, t^{(1)}_{11}]=t^{(k)}_{12},&[t^{(1)}_{12},
t^{(k)}_{22}]=0,\\ {[t^{(1)}_{11},t^{(k)}_{21}]}=t^{(k)}_{21}, &[t^{(1)}_{21},
t^{(k)}_{22}]=0,\ (k=1,2)\\ {[t^{(1)}_{12},t^{(k)}_{21}]}=t^{(k)}_{22},&
t^{(1)}_{22}\ \mbox{is a center}.\end{array}$ \hfill (\ref{p5.2}')\\

\noindent Because $t^{(1)}_{22}$ is a center and
\begin{equation}
C_2=t^{(2)}_{22}+t^{(1)}_{22}+t^{(1)}_{11}t^{(1)}_{22}-t^{(1)}_{12}
t^{(1)}_{21} \label{p5.5}
\end{equation}
is then also a center:
\begin{equation}
[C_2,t^{(k)}_{ab}]=0.\label{p5.6}
\end{equation}
One is able to show that the relation given by eq.~(\ref{p5.3}) can be
reduced to\\

\hspace*{0.6cm}
$[t^{(2)}_{22},t^{(k)}_{11}]+t^{(k)}_{21}t^{(1)}_{12}-t^{(1)}_{21}t^{(k)}_{12}
=0,\ (k\geq 2).$\hfill (\ref{p5.3}')\\

\noindent Therefore now eqs.~(\ref{p5.1}), (\ref{p5.2}'), (\ref{p5.3}') and
(\ref{p5.4})
are independent relations. The simplest realization of eq.~(\ref{p5.2}')
for $k=1$ is
\begin{equation}
t^{(1)}=\left[\begin{array}{cc}M&\bar{P}\\P&C\end{array}\right]\label{p5.7}
\end{equation}
where $C$ is a center and
\begin{equation}
[M,P]=P,\ \ [M, \bar{P}]=-\bar{P},\ \ [P,\bar{P}]=C\label{p5.8}
\end{equation}
A simple realization of eq.~(\ref{p5.7}) is when $C=0$:
\begin{equation}
t^{(1)}=\left[\begin{array}{cc}ip&e^{-q}\\e^q&0\end{array}\right]\label{p5.9}
\end{equation}
where
\begin{equation}
[p,q]=-i\label{p5.10}
\end{equation}
i.e., $p$ and $q$ obey the Heisenberg algebra.

To determine $t^{(2)}$ we should find a realization satisfying
eqs.~(\ref{p5.1})
, (\ref{p5.2}') and (\ref{p5.3}') for $k=2$. The corresponding relations for
$k>2$ yield the constraints to the considered operators. This will gives
rise to a new type of infinite algebra. However what we are interested in
is to find a more physical example where the expansion of $t(u)$ is truncated
\begin{equation} t(u)=\sum^m_{n=0}u^{-n}t^{(n)}\label{p5.11} \end{equation}
i.e.,
\begin{equation} t^{(m+1)}=0,\ \ (\mbox{all}\ t^{(k)}=0\ \mbox{for}\
k\geq m+1). \label{p5.12} \end{equation}
This possibility is prohibited for the case (A) which leads to {\em Yangian},
however,
eq.~(\ref{p5.12}) can be allowed by $t^{(0)}=\left[\begin{array}{cc}1&0\\
0&0\end{array}\right]$. This fact can be verified by direct calculation.
Let us look for an exact solution of $m=3$, i.e., by substituting
\begin{eqnarray}
t(u)=\left[\begin{array}{cc}1&0\\0&0\end{array}\right]+u^{-1}t^{(1)}
+u^{-2}t^{(2)}+u^{-3}t^{(3)}\label{p5.13}
\end{eqnarray}
into the RTT relation to find a sufficient solution of $t^{(1)},\ t^{(2)}$
and $t^{(3)}$. This solution is related to the Goryachev-Chaplegin
gyrostat discussed in Refs.~\cite{r8,r6}.

This model is interesting for the finite terms in the expansion of $t(u)$.
The truncation of $t(u)$ corresponds to the finite number of conserved
quantities and the satisfaction of RTT relation indicates the quantum
integrability of the model.

The Goryachev-Chaplygin(G-C) top is completely integrable in sense of
Liouville as shown in the literature~\cite{r8}. It describe an axially
symmetric top moving in an applied field. In the quantum case the G-C
top is extended to the quantum G-C gyrostat whose Hamiltonian takes the
form:
\begin{equation}
H=\frac 12 (J^2+3J^2_3)-bx_1+pJ_3\label{e1}
\end{equation}
where $J^2=J^2_1+J^2_2+J^2_3$ and $b$ a parameter. We should note that in this
section the notations $J_i$ express the angular momentum. They are
irrelevant to the $J_{\lambda}$-operators for {\em Yangian} shown before.
The quantities appearing in eq.~(\ref{e1}) satisfy the following commutation
relations:
\begin{eqnarray}
[J_i, J_j]   &=& -i\epsilon_{ijk}J_k\ ,\nonumber\\
{[J_i,x_j]}  &=& -i\epsilon_{ijk}x_k\ ,\nonumber\\
{[x_i,x_j]}  &=& 0\ ,\label{e2}\\
{[p,q]}      &=& -i\ ,\nonumber\\
{[p,J_i]}     &=& [p,x_i]=[q,x_i]=[q,J_i]=0\nonumber
\end{eqnarray}
where $i,\ j,\ k$ take over $1,\ 2$ and $3$. The momentum $p$ and coordinate
$q$ of mass center commute with all the locally dynamic variables.
The Hamiltonian eq.~(\ref{e1}) commutes with the other integral of motion:
\begin{equation}
G=(2J_3+p)(J^2-J^2_3)+2b[x_3,J_1]_+\ .
\end{equation}
where $[A,B]_+\equiv AB+BA$.
In addition to the commutation relations eq.~(\ref{e2}) there are two
constraints:
\begin{eqnarray}
\sum^3_{i=1}J_ix_i &=& 0 \label{e4},\\
\sum^3_ix^2_i &=& 1  \label{e5}.
\end{eqnarray}
Eq.~(\ref{e4}) plays important roles in proving the quantum integrability of
the model.

Now let us turn to the RTT relation.

For solving eq.~(\ref{p1.7}) with the form eq.~(\ref{p5.13}) one first put
\begin{equation}
t^{(1)}_{11} = \alpha p, \ \ t^{(1)}_{22} =0,\ \ t^{(1)}_{12}=
  \beta e^{\tau q}x_+ ,\ \  t^{(1)}_{21} =\gamma e^{-\tau q},\label{e17}
\end{equation}
and
\begin{eqnarray}
t^{(2)}_{11} &=& f_1J^2+f_2J^2_3+f_3pJ_3+f_4x_-+f_5,\nonumber\\
t^{(3)}_{11} &=& (p+g_3J_3)(g_1J^2+g_2J^2_3+g_5)+g_4[J_-,x_3]_+\label{e18}
\end{eqnarray}
where $f_1,\ldots,f_5,\ g_1,\ldots,g_4,\ \alpha,\ \beta,\ \gamma$ and $\tau$
all parameters to be determined and $J_{\pm}=J_1\pm iJ_2$, $x_\pm=x_1
\pm ix_2$ obey the commutation relations shown by eq.~(\ref{e2}).
Obviously not all the parameters are independent.
$\lambda$ can be normalized to be one. Substituting
eqs.~(\ref{e17}) and (\ref{e18}) together with eq.~(\ref{e4}) into
eq.~(\ref{p1.7}) after lengthy calculations by hand we find
\begin{eqnarray*}
\lefteqn{ \tau=-i\alpha^{-1}\lambda,\ f_1=-\frac 14\lambda,
   \ f_2=-\frac 34\lambda,\ f_3=\alpha,\ f_5=-\frac 1{16}\lambda,} \\
\lefteqn{ g_1=-g_2=-\frac 14 \alpha,\ g_3=-\lambda\alpha^{-1},
   \ g_4=\frac 14f_4, \ g_5=-\frac 1{16}\alpha.}
\end{eqnarray*}
Denoting $f_4=f$ the solution of $t^{(n)}$ reads:
\begin{eqnarray}
t^{(2)}_{11} &=& -\frac 14 \lambda (J^2+3J^2_3+\frac 14)+\alpha p J_3+f x_-,
\ \ t^{(2)}_{22}= \lambda^{-1}\gamma \beta x_+,\nonumber\\
t^{(2)}_{12} &=& - \lambda^{-1}\beta e^{\tau q}\left( -\frac{\lambda}4
  [J_+, x_3]_+ +x_+(\lambda J_3-\alpha p)\right),\ \ t^{(2)}_{21} =
  \gamma e^{-\tau q}J_3,\\
t^{(3)}_{11} &=& -\frac 14 \alpha (p-\lambda \alpha^{-1}J_3)(J^2-J^2_3+
  \frac 14) +\frac 14 f[J_-, x_3]_+,\nonumber\\
t^{(3)}_{22} &=& \frac 14 \lambda^{-1}\beta \gamma [J_+, x_3]_+,\nonumber\\
t^{(3)}_{12} &=& -\lambda^2\beta e^{\tau q}\left\{ fx^2_3-\frac 14 \alpha
[J_+, x_3]_+(p-\lambda \alpha^{-1}J_3)\right\},\nonumber\\
t^{(3)}_{21} &=& -\frac 14\gamma e^{-\tau q}(J^2-J^2_3+\frac 14)\label{e25}.
\end{eqnarray}
Therefore the truncated {\em Yangian} can be viewed as a mapping of the algebra
given by eq.~(\ref{e2}).

Substituting the derived $t(u)$ into eq.~(\ref{p2.54}) it follows
\begin{equation}
\mbox{det}t(u)=u^{-3}(u-1)^{-3}\lambda^{-1}f(u^2-u+\frac 3{16})
  (x_+x_-+x^2_3).
\end{equation}
So only $\sum^3_{i=1}x^2_i$ is a Casimir operator, i.e. $C_m=0$ for $m<4$,
$C_4=\lambda^{-1}f\sum^3_{i=1}x^2_i$ that commute with $J_i$. The
constraint eq.~(\ref{e5}) does not play role in solving the RTT relation.
By choosing a proper representation eq.~(\ref{e5}) can be taken that
is automatically satisfied by virtue of eq.~(\ref{p2.48}).
Notice that det$t(u)$ has zero's at $u=\frac 14$ and
$u=\frac 34$ where the inverse of $t(u)$ can not be defined.

It is worth noting that the commutation relations shown by eq.~(\ref{e2})
are invariant subject to a transformation:
\begin{equation}
J^{\prime}_a=\sum_b A_{ab}J_b,\ x^{\prime}_a=\sum_b A_{ab}x_b,
\ \ (a,b=1,2)\label{e36}
\end{equation}
where
\begin{eqnarray}
\begin{array}{ll} A_{11}=\epsilon A_{22}, &A_{12}=-\epsilon A_{21},\\
A^2_{11}+A^2_{21}=1, & \epsilon=\pm 1=\mbox{det}A \end{array}.
\end{eqnarray}
This transformation is useful to transform the Hamiltonian in preserving the
RTT relation.

Let us turn to the conserved quantities for G-C gyrostat.

By taking the trace of the transfer matrix one obtains
\begin{eqnarray}
\hspace{-1.2cm}
\mbox{tr}t(u)&=&\lambda+u^{-1}\alpha p+u^{-2}\left\{-\frac 14 \lambda
   (J^2+3J^2_3+\frac 14)+\alpha p J_3+fx_-+\lambda^{-1}\beta\gamma x_+
   \right\}+\nonumber\\
&&\frac {u^{-3}}4\left\{-(\alpha p-\lambda J_3)(J^2-J^2_3+
   \frac 14)+ f[J_-,x_3]_++\lambda^{-1}\beta\gamma
   [J_+,x_3]_+\right\}.
\end{eqnarray}
Hence we have the Hamiltonian $H_p$ and the other conserved quantities $G_p$:
\begin{eqnarray}
\lefteqn{H_p=\frac 12\left\{\frac 14\lambda (J^2+3J^2_3)-fx_--\lambda^{-1}
\beta\gamma x_+-\alpha p J_3 \right\}+\frac 1{16}\lambda \label{e29},}\\
\lefteqn{G_p=\frac 14 \left\{(\alpha p-\lambda J_3)(J^2-J^2_3+\frac 14)+
   f[J_-,x_3]_++\lambda^{-1}\beta\gamma[J_+,x_3]_+\right\}\label{e30}.}
\end{eqnarray}
If one requires
\begin{equation}
4\lambda^{-1}\beta\gamma f=b^2,\label{e31}
\end{equation}
then there is rotational invariance about the $x_3$ axis. By virtue of
eqs.~(\ref{e25}) and (\ref{e36}) we find that eqs.~(\ref{e29}) and (\ref{e30})
are reduced to
\begin{eqnarray}
H_p &=& \frac 12\left\{\frac 14\lambda ({J^{\prime}}^2+3{J^{\prime}}^2_3)
   -b{x^{\prime}}_1-\alpha p\epsilon {J^{\prime}}_3\right\}+
   \frac 1{16}\lambda,\\
G_p &=& -\frac 14 \left\{(\alpha p-\epsilon\lambda{J^{\prime}}_3)
   ({J^{\prime}}^2-{J^{\prime}}^2_3+\frac 14)+b[x_3,{J^{\prime}}_1]_+
   \right\}
\end{eqnarray}
which are exactly those given by Sklyanin~\cite{r6}. Obviously,
the parameters $\lambda$ and $\tau$ are trivial in the solution of
$t(u)$ and $\beta$, $\gamma$ can be viewed as the consequence of
$t(u)$ subject to a similar transformation. Therefore only parameters
$f$ and $\alpha$ are essential in determining the form of $t(u)$
though $f$ does not appear in the Hamiltonian. The simplest realization of
quantum G-C top is a quantum mechanical system on sphere.

Taking the constraints eqs.~(\ref{e4}) and (\ref{e5}) into account the
simplest realization of $J_i$ and $x_i$ can be made through
\begin{eqnarray}
x_i &=& n_i(t),\\
J_i &=& -i\epsilon_{ijk}n_j(t)\dot n_k(t)\ (i,j,k=1,2,3)
\end{eqnarray}
that give $J^2=-\dot{\vec n}\cdot\dot{\vec n}$.

In conclusion we have shown that the truncated {\em Yangian} gives rise to
quantum integrable system with finite number of conserved quantities and
make the system work in the space higher than (1+1) dimensions.

The trigonometric extension of the truncated {\em Yangian} will be the
``truncated'' affine quantum algebra which is the generalization of
Drinfeld's statement discussed in Ref.~\cite{r3,r5}. It gives
the $q$-deformed G-C gyrostat which will be discussed in the next section.

\section{Trigonometric G-C Gyrostat}
\setcounter{equation}{0}

If $R$-matrix is taken to be the simple 6-vertex from~(\ref{p1.20}), the
trigonometric
transfer matrix  $T(x)$ can be found on the basis of RTT relation, where
$x=e^u$(or $x=e^{iu}$) is the spectral parameter. In terms of $x$, the
RTT relation takes the form
\begin{equation}
{\check R}(xy^{-1})(T(x)\otimes T(y))=
(T(y)\otimes T(x)){\check R}(xy^{-1}) \label{p6.1}
\end{equation}
$T(x)$ can be expanded as
\begin{equation}
T(x)=\sum^{+\infty}_{n=-\infty}x^nT^{(n)}.\label{p6.2}\end{equation}
A simple form of trigonometric $R$-matrix(which is a deformation of
eq.~(\ref{p1.21})) is given by
\begin{eqnarray}
\check{R}(u)=\left[\begin{array}{cccc} \sin (u+\eta) & & & \\ & e^{-i\theta}
\sin \eta & \sin u& \\ &\sin u& e^{i\theta}\sin \eta & \\ & & &\sin (u+\eta)
\end{array}\right] \label{p6.3}
\end{eqnarray}
where $x=e^{iu}$, $\eta$ and $\theta$ are constants. Set $q=e^{-i\eta}$,
eq.~(\ref{p6.3}) can be recast into
\begin{equation}
{\check R}(x)=x{\check R}-x^{-1}{\check R}^{-1}\label{p6.4}
\end{equation}
where
\begin{eqnarray}
\check{R}&=&\left[\begin{array}{cccc}q^{-1}& & & \\
&0&1& \\ &1&q^{-1}-q& \\ & & &q^{-1}\end{array}
\right]\label{p6.5} \end{eqnarray}
which is the asymptotic form of ${\check R}(x)$ as $x\rightarrow \infty$.
It can be verified that there is a subset in $T(x)$:
\begin{equation}   T(x)=xT_+-x^{-1}T_- \label{p6.6}   \end{equation}
satisfying eq.~(\ref{p6.2}) for the ${\check R}$ given by eq.~(\ref{p6.3}).
Actually, substituting eq.~(\ref{p6.5}) into eq.~(\ref{p6.3})
it follows~\cite{r1}:
\begin{eqnarray}\begin{array}{l}
{\check R}(T_{\pm}\otimes T_{\pm})=(T_{\pm}\otimes T_{\pm}){\check R}\\
{\check R}(T_{+}\otimes T_{-})=(T_{-}\otimes T_{+}){\check R}
\end{array}\label{p6.7}  \end{eqnarray}
where $\check R$ is given by eq.~(\ref{p6.5}). It is noted that each element
of the matrices $T_{\pm}$ is quantum operator. One can check that
\begin{eqnarray}
T_+=\left[\begin{array}{cc}K&(q^{-1}-q)X^+\\0&K^{-1}\end{array}\right],\ \
T_-=\left[\begin{array}{cc}K^{-1}&0\\(q^{-1}-q)X^-&K\end{array}\right]
\label{p6.8}     \end{eqnarray}
satisfy eq.~(\ref{p6.7}) when $X^{\pm}$ and $K$ satisfy
\begin{equation}
KX^{\pm}K^{-1}=q^{\pm}X^{\pm},\ \ {[X^+,X^-]}=\frac{K^2-K^{-2}}{q-q^{-1}}
\label{p6.9} \end{equation}
that is known as the quantum algebra associated with $SU(2)$, denoted by
$U_q(SU(2))$. The pioneered works concerning quantum algebras were made by
Drinfeld~\cite{r3,r5}, Jimbo~\cite{r11} and Faddeev~\cite{r9}
although the simplest form eq.~(\ref{p6.9}) was first presented
by Kulish and Sklyanin~\cite{re5}.

The general form of trigonometric $T(x)$ is very complicated and denoted by
$U_q(\widehat{SU(2)})$, i.e., the $q$-affine algebra~\cite{r3}. The comparison
between $Y(sl(2))$ and $U_q(sl(2))$ is shown by the following chart-follow
($w=q-q^{-1}$):
\vspace*{0.5cm}

\noindent\makebox[2in]{${\check R}=uP+I$}
\makebox[0.5in]{$\stackrel{q=e^{-i\eta}}{\longleftarrow}$}
\makebox[2.5in]{${\check R}$ given by eq.~(\ref{p6.3})}\\
%%%%%%%%%%%%%%%%
\makebox[2in]{$\downarrow$}
\makebox[0.5in]{$\begin{array}{c}\mbox{\small rational}\\
\mbox{\small limit}\end{array}$}
\makebox[2.5in]{$\downarrow$}\\
%%%%%%%%%%%%%%%
\makebox[0.3in]{} \framebox[4.2in]{RTT relation}\\
%%%%%%%%%%%%%%%%
\makebox[2in]{$\downarrow$} \makebox[0.5in]{}
\makebox[2.5in]{$\downarrow$}\\
%%%%%%%%%%%%%%%%%%%%
\makebox[2in]{}
\makebox[0.5in]{}
\makebox[2.5in]{$T(x)=\sum^{+\infty}_{n=-\infty}x^nL^{(n)}$}\\
%%%%%%%%%%%%%%%
\makebox[2in]{$T(u)=\sum^{\infty}_{n=0}u^{-n}T^{(n)}$}
\makebox[0.5in]{}
\makebox[2.5in]{$=L^{(0)}+xL^{(1)}+x^{-1}L^{(-1)}+\cdots
%\sum_{|n|\geq 2}L^{(n)}
$}\\
%%%%%%%%%%%%%%%%%%%%%
\makebox[2.5in]{}
\makebox[2.5in]{$L^{(1)}=L_+,\ L^{(-1)}=-L_-$}\\
\makebox[2in]{$T^{(1)}=\left[\begin{array}{cc}I_3&I_+\\I_-&-I_3\end{array}\right]$}
\makebox[0.5in]{$\begin{array}{c}\mbox{{\small rational}}\\
\longleftarrow \\ \mbox{{\small limit}}\end{array}$}
\makebox[2.5in]{$\begin{array}{c}
L_+=\left[\begin{array}{cc}K&wX^+\\0&K^-\end{array}\right],\\
L_-=\left[\begin{array}{cc}K^{-1}&0\\-wX^-&K\end{array}\right]
\end{array}$}\\
%%%%%%%%%%%%%%%%
\makebox[2in]{$J_{\lambda},\ T^{(n)}(n\geq 3)$}
\makebox[0.5in]{}
\makebox[2.5in]{$\sum_{n}x^nL^{(n)}$}\\
%%%%%%%%%%%%%%%
\makebox[2in]{$Y(sl(2))$}
\makebox[0.5in]{}
\makebox[2.5in]{$U_q(sl(2))$}\\
%%%%%%%%%%%%%%%
\makebox[2in]{deformed loop algebra}
\makebox[0.8in]{}
\makebox[2.2in]{$q$-deformed Kac-Moody algebra}
\vspace*{0.5cm}

To find such $T(x)$-matrix let us consider 2 by 2 matrix, substitute
\begin{eqnarray}
T(x)=\left[\begin{array}{cc}T_{11}(x)&T_{12}(x)\\T_{21}(x)&T_{22}(x)
 \end{array}\right]=\sum^{+\infty}_{n=-\infty}x^nT^{(n)} \label{f1}
\end{eqnarray}
or
\begin{equation}
T_{ab}(x)=\sum^{+\infty}_{n=-\infty}x^nT^{(n)}_{ab}\ (a,b=1,2) \label{f2}
\end{equation}
into eq.~(\ref{p6.1}) we derive
\begin{eqnarray}
\lefteqn{[T^{(k)}_{ab}, T^{(j)}_{ab}]=0,\ \ \
[T^{(k)}_{11}, T^{(j)}_{22}]=[T^{(j)}_{11}, T^{(k)}_{22}], \ \ \
[T^{(k)}_{12}, T^{(j)}_{21}]=[T^{(j)}_{12}, T^{(k)}_{21}],\nonumber}\\
\lefteqn{{[T^{(k-1)}_{22},T^{(j+1)}_{11}]}-[T^{(k+1)}_{22},T^{(j-1)}_{11}] =
w(T^{(j)}_{12}T^{(k)}_{21}-T^{(k)}_{12}T^{(j)}_{21}), \nonumber}\\
\lefteqn{{[T^{(k-1)}_{21},T^{(j+1)}_{12}]}-[T^{(k+1)}_{21},T^{(j-1)}_{12}] =
w(T^{(j)}_{11}T^{(k)}_{22}-T^{(k)}_{11}T^{(j)}_{22}, \label{f3}}\\
\lefteqn{qT^{(k-1)}_{aa}T^{(j+1)}_{ab}-q^{-1}T^{(k+1)}_{aa}T^{(j-1)}_{ab} =
T^{(j+1)}_{ab}T^{(k-1)}_{aa}-T^{(j-1)}_{ab}T^{(k+1)}_{aa}
+w T^{(j)}_{aa}T^{(k)}_{ab}, \nonumber}  \\
\lefteqn{qT^{(k-1)}_{ab}T^{(j+1)}_{aa}-q^{-1}T^{(k+1)}_{ab}T^{(j-1)}_{aa} =
T^{(j+1)}_{aa}T^{(k-1)}_{ab}-T^{(j-1)}_{aa}T^{(k+1)}_{ab}
+w T^{(j)}_{ab}T^{(k)}_{aa}, \nonumber}\\
\lefteqn{qT^{(j+1)}_{ba}T^{(k-1)}_{aa}-q^{-1}T^{(j-1)}_{ba}T^{(k+1)}_{aa} =
T^{(k-1)}_{aa}T^{(j+1)}_{ba}-T^{(k+1)}_{aa}T^{(j-1)}_{ba}
+w T^{(k)}_{ba}T^{(j)}_{aa}, \nonumber}\\
\lefteqn{qT^{(j+1)}_{aa}T^{(k-1)}_{ba}-q^{-1}T^{(j-1)}_{aa}T^{(k+1)}_{ba} =
T^{(k-1)}_{ba}T^{(j+1)}_{aa}-T^{(k+1)}_{ba}T^{(j-1)}_{aa}
+w T^{(j)}_{aa}T^{(k)}_{ba}. \nonumber}
\end{eqnarray}
The inverse of $T(x)$ is given by
\begin{eqnarray}
[T(x)]^{-1}=[\mbox{det}_qT(x)]^{-1}\left[\begin{array}{cc}T_{22}(q^{-1}x)&
-T_{12}(q^{-1}x)\\-T_{21}(q^{-1}x)&T_{11}(q^{-1}x)\end{array}\right] \label{f4}
\end{eqnarray}
where
\begin{equation}
\mbox{det}_q T(x)=T_{11}(x)T_{22}(q^{-1}x)-T_{12}(x)T_{21}(q^{-1}x) \label{f5}
\end{equation}
It is easy to know that
\begin{equation}
[\mbox{det}_qT(x), T_{ab}(y)]=0 \label{f6}
\end{equation}
Correspondingly
\begin{equation}
\mbox{det}_qT(x)= \sum_{n=-\infty}^{+\infty}x^{-n}C_n, \ \
C_n=\sum_{k+j=n}q^{-j}(T^{(k)}_{11}T^{(j)}_{22}-T^{(k)}_{12}T^{(j)}_{21})
\label{f7}
\end{equation}
and
\begin{equation}
\mbox{tr} T(x)=\sum_{n=-\infty}^{+\infty}(T^{(n)}_{11}+T^{(n)}_{22})x^n
\label{f8}
\end{equation}

\noindent {\bf Definition} $T^{(n)}$ is called truncated $q$-affine quantum
algebra
if $T^{(n)}=0$ for $|n|\geq 4$.

To solve (\ref{f3}) we take the ansatz
(motivated by the rational correspondence)
\begin{eqnarray}
T^{(\pm 2)}_{11}&=&T^{(0)}_{11}=T^{(\pm 3)}_{22}=T^{(\pm 2)}_{22}=
T^{(0)}_{22}\nonumber\\
&=&T^{(\pm 3)}_{12}=T^{(\pm 1)}_{12}=T^{(\pm 3)}_{21}=T^{(\pm
1)}_{21}=0\label{f9}
\end{eqnarray}
and set
\begin{eqnarray}
\lefteqn{T^{(\pm 3)}_{11} =\pm\lambda_3e^{\mp i\eta P},\ \ \ \
T^{(\pm 1)}_{11}=A^{(0)}_{\pm}+e^{\mp i\eta P}A^{(1)}_{\pm}+e^{\pm i \eta P}
A^{(2)}_{\pm}}\\
\lefteqn{T^{(\pm 2)}_{12}=e^{i\xi Q\mp i\eta P}E_{\pm 2}, \ \ T^{(0)}_{12}=
 e^{i\xi Q}(E^{(0)}+e^{-i\eta P}E^{(-)}+e^{i\eta P}E^{(+)}) \label{f11}}\\
\lefteqn{T^{(\pm 2)}_{21}=\alpha_2K^{\pm 1}e^{-i\xi Q},\ \
T^{(0)}_{21}=e^{-i\xi Q}F_0,\ \ q=e^{i\xi\eta}}
\end{eqnarray}
where $\lambda_3,\ \alpha_2, \ \xi$ and $\eta$ are constants, $P$ and $Q$
satisfy
\begin{equation}
[P,Q]=-i \label{f12}
\end{equation}
here $A^{(i)}_{\pm}(i=0,1,2)$, $E_{\pm 2},\ E^{(j)}(j=0,\pm),\ F_0$ and $K$ are
operators commuting with $P$ and $Q$. They will be determined by (\ref{f3}).

After tedious calculation we find that the following algebraic relations solve
the RTT relation for 6-vertex form of $R$-matrix:
\begin{eqnarray}
\lefteqn{KE_{\pm 2}=q^{-1}E_{\pm 2}K \label{f13}\nonumber}\\
\lefteqn{A^{(2)}_{\pm}=\mp \lambda_3 K^{\pm 2},\ \ \ A^{(1)}_{\pm}=\pm
\lambda_3
 \alpha^{-1}_2K^{\mp}S,\ \ F_0=S\label{f14}}\\
\lefteqn{E^{\pm}=-E_{\pm 2}K^{\pm 2}, \ \ T^{(\pm 1)}_{22}=\pm
\lambda^{-1}_3\alpha_2
E_{\pm 2}K^{\pm 1} \nonumber}
\end{eqnarray}
where $A^{(0)}_{\pm},\ E_{\pm 2},\ E^{(0)}$ and $S$ satisfy
\begin{eqnarray}
\lefteqn{q^2E_{-2}E_{+2}=E_{+2}E_{-2},\ \ [A^{(0)}_+,
A^{(0)}_-]=[K,S]=0\nonumber}\\
\lefteqn{KA^{(0)}_{\pm}K^{-1}=q A^{(0)}_{\pm},\ \ q^{\pm 1}E_{\pm
2}A^{(0)}_{\mp}
=A^{(0)}_{\mp}E_{\pm 2}\nonumber}\\
\lefteqn{q^{\pm 1}E_{\pm 2}A^{(0)}_{\pm}-A^{(0)}_{\pm}E_{\pm 2}=\lambda_3w
E^{(0)}
\label{f15}}\\
\lefteqn{SA^{(0)}_{\pm} -q^{\mp 1}A^{(0)}_{\pm} S = \pm \alpha_2 w
A^{(0)}_{\mp}
K^{\pm 1} \nonumber}\\
\lefteqn{SE_{\pm 2} -q^{\pm 1}E_{\pm 2} S=\pm \alpha_2 w E_{\mp 2}K^{\mp
1}\nonumber}
\end{eqnarray}

A realization of the algebra eqs.~(\ref{p6.2}) and (\ref{f15}), i.e.,
$T^{(n)}$($T^{(n)}=0$ for $|n|\geq 4$), can be made by
\begin{eqnarray}
\lefteqn{A^{(0)}_{\pm}=\lambda^{(1)}_{\pm}{\hat J}_-{\hat
x}_3+\lambda^{(2)}_{\pm}
{\hat x}_-,\nonumber}\\
\lefteqn{E_{\pm 2}=\beta^{(1)}_{\pm}{\hat J}_+{\hat x}_3+\beta^{(2)}_{\pm}
{\hat x}_+,\nonumber}\\
\lefteqn{S=\lambda_1{\hat J}_+{\hat J}_-+\lambda_2\label{g1},}\\
\lefteqn{K=\exp{(i\xi\eta {\hat J}_3)}\nonumber}
\end{eqnarray}
where $E^{(0)},\ \lambda^{(i)}_{\pm},\ \beta^{(i)}_{\pm}$ and
$\lambda_i(i=1,2)$
are $K$-dependent operators which are given by
\begin{equation}
\alpha^2_2=g\alpha^2,\ \ \lambda_1=(g-1)wg^{-1}\alpha,\
\lambda_2=\alpha(q-K+K^{-1}), \nonumber\\
\end{equation}
and
\begin{eqnarray}
\lambda^{(1)}_-&=&0,\ \lambda^{(1)}_{+}=\lambda^{(+)}K^{2(1-\delta_+)-\beta},
\ \lambda^{(2)}_{+}=(1-q^{-1})^{-1}q^{-\delta_+}\tau_-K^{-1+\delta_+}
\lambda^{(1)}_{+}\nonumber\\
\lambda^{(2)}_{-}&=&(1-q^{-1})^{-1}q^{-1-\delta_+}\alpha^{-1}\alpha_2
\tau_-K^{-1+\delta_+}\lambda^{(1)}_{+},\ \ \beta^{(1)}_{+}=0,\nonumber\\
\beta^{(2)}_{-}&=&-(1-q^{-1})^{-1}q^{-1+\delta_+}\tau_+K^{\delta_+}
\beta^{(1)}_{-},\ \ \beta^{(1)}_{-}=\beta^{(-)}K^{\beta}\label{g4},\\
\beta^{(2)}_{+}&=&(1-q^{-1})^{-1}q^{-4+\delta_+}\alpha^{-1}
\alpha_2\tau_+K^{\delta_+ -2}\beta^{(1)}_{-},
\nonumber\\
E^{(0)}&=&-\lambda^{-1}_3w^{-1}(1-q^{-1})^{-1}q^{-3+\delta_+}g
\alpha^{-1}\alpha_2K^{- 1}\beta^{(1)}_{-}
\lambda^{(1)}_{+}(qK)({\hat x}_3)^2\nonumber
\end{eqnarray}
for $\delta_++\delta_-=1$.
\begin{eqnarray}
\lambda^{(1)}_+&=&0,\ \ \lambda^{(1)}_{(-)}=\lambda^{-}K^{2(1+\delta_+)-\beta},
\ \lambda^{(2)}_{-}=(1-q^{-1})^{-1}q^{-\delta_+}\tau_-K^{1+\delta_+}
\lambda^{(1)}_{-},\nonumber\\
\lambda^{(2)}_{+}&=&(1-q^{-1})^{-1}q^{-\delta_+}\alpha^{-1}\alpha_2
\tau_-K^{1+\delta_+}\lambda^{(1)}_{-},\ \ \beta^{(1)}_{-}=0,\nonumber\\
\beta^{(2)}_{+}&=&-(1-q^{-1})^{-1}q^{1+\delta_+}\tau_+K^{\delta_+}
\beta^{(1)}_{+},\ \ \beta^{(1)}_{+}=\beta^{(+)}K^{\beta}\label{g4'},\\
\beta^{(2)}_{-}&=&(1-q^{-1})^{-1}q^{4+\delta_+}\alpha
\alpha^{-1}_2\tau_+K^{\delta_++ 2}\beta^{(1)}_{+},
\nonumber\\
E^{(0)}&=&-\lambda^{-1}_3w^{-1}(1-q^{-1})^{-1}q^{3+\delta_+}g
\alpha\alpha^{-1}_2K\beta^{(1)}_{+}
\lambda^{(1)}_{-}(qK)({\hat x}_3)^2\nonumber
\end{eqnarray}
for $\delta_++\delta_-=-1$, where $\delta_+,\ \alpha,\ \beta^{(\pm)},\
\lambda^{(\pm)}$ and $\tau_{\pm}$ are arbitrary constants.
It is noted that the parameters $\beta^{(1)}_{\mp}(K)$ depend on $K$
and $\lambda^{(1)}_{\pm}(qK)$ on $qK$, respectively, whereas $\delta_+,\
\alpha,\
\beta^{(\pm)},\ \lambda^{(\pm)}$ and $\tau_{\pm}$ are arbitrary constants.

The operators ${\hat J}_{\pm},\ {\hat J}_{3},\  {\hat x}_{\pm}$ and
${\hat x}_{3}$ in eq.~(\ref{g1}) obey the following relations
\begin{eqnarray}
[{\hat J}_{\pm},{\hat x}_{\pm}] &=& [{\hat J}_3,{\hat x}_3] =0,\nonumber\\
{[{\hat J}_{\pm}, {\hat J}_3]} &=& \pm {\hat J}_3,\ \
[{\hat J}_+, {\hat J}_-] =-g[{\hat J}_3]_q,\nonumber\\
{[{\hat x}_{\pm},{\hat J}_3]} &=& \pm {\hat x}_{\pm},\nonumber\\
q^{\delta_{\mp}}{\hat J}_{\pm}{\hat x}_3 &=& {\hat x}_3{\hat J}_{\pm}
\pm \tau_{\pm}K^{\pm\delta_{\pm}}{\hat x}_{\pm}\label{f18},\\
{[{\hat x}_{+},{\hat x}_{-}]} &=& 0,\ \ \
{\hat x}_{\pm}{\hat x}_{3}=q^{\delta_{\pm}}{\hat x}_3{\hat x}_{\pm},\nonumber\\
q^{-1}{\hat J}_{\pm}{\hat x}_{\mp} &=& {\hat x}_{\mp}{\hat J}_{\pm}
\mp \tau^{-1}_{\mp}gK^{\mp\delta_{\pm}}{\hat x}_{3}\nonumber
\end{eqnarray}
together with the constraint
\begin{equation}
g[{\hat J}_3]_q{\hat x}_3+\tau_-K^{-\delta_-}{\hat J}_+{\hat x}_-+\tau_+
K^{\delta_+}{\hat J}_-{\hat x}_+=0\label{f19}
\end{equation}

In conclusion eqs.~(\ref{f11})-(\ref{f19}) solve RTT relation for $R$ is
given by eq.~(\ref{p1.18}).

The $\mbox{det}_qT(x)$ for $T^{(n)}(x)=0(|n|\geq 4)$ can
be expressed by ${\hat J}_{\pm},\ {\hat x}_{\pm}$ and ${\hat x}_3$ through
\begin{equation}
\mbox{det}_qT(x)=\{x^2+q^2x^{-2}+\alpha^{-1}\alpha_2(1+q)\}C_{+2}
\ \ (\alpha^{-1}\alpha_2=\pm q^{\frac 12}) \label{g5}
\end{equation}
where
\begin{equation}
C_{+2}=F_2\{\tau_-K^{\delta_+}({\hat J}_+{\hat J}_-{\hat x}_3-\frac{\tau_+
K^{\delta_+}{\hat x}_+{\hat x}_-}{1-q}+w^{-1}g({\hat x}_3)^2)\},\
(w=q-q^{-1})\label{g6}
\end{equation}
with
\begin{equation}
F_2=\lambda^{-1}_3\alpha(1-q)^{-1}q^{1-\beta-\delta_+}\beta^{(+)}
\lambda^{(-)}K^{2\delta_+}\label{g7}
\end{equation}
that leads to for $\delta_+ + \delta_-=-1$:
\begin{equation}
C_{+}=F_1\{\tau_+K^{\delta_+-1}({\hat J}_-{\hat J}_+{\hat x}_3-\frac{\tau_-
K^{\delta_+-1}{\hat x}_+{\hat x}_-}{1-q^{-1}}+wg({\hat x}_3)^2)\}
\end{equation}
where
\begin{equation}
F_1=\lambda^{-1}_3\alpha(1-q^{-1})^{-1}q^{-2-\beta-\delta_+}\beta^{(-)}
\lambda^{(+)}K^{2(1-\delta_+)}
\end{equation}
The non-vanishing $C_n$ in (\ref{f7}) are
\begin{eqnarray}
\lefteqn{C_{\pm 4} = q^{\mp 1}(T^{(\pm 3)}_{11}T^{(\pm 1)}_{22}
 -q^{\mp 1}T^{(\pm 2)}_{12}T^{(\pm 2)}_{21}),\nonumber}\\
\lefteqn{C_{\pm 2} = q^{\pm 1}T^{(\pm 3)}_{11}T^{(\mp 1)}_{22}+q^{\mp 1}
T^{(\pm 1)}_{11}T^{(\pm 1)}_{22}-T^{(\pm 2)}_{12}T^{(0)}_{21}-
q^{\mp 2}T^{(0)}_{12}T^{(\pm 2)}_{21},\label{g8}}\\
\lefteqn{C_{0} = qT^{(+1)}_{11}T^{(-1)}_{22}+q^{-1}T^{(-1)}_{11}T^{(+1)}_{22}
-q^2T^{(2)}_{12}T^{(-2)}_{21}-q^{-2}T^{(-2)}_{12}T^{(2)}_{21}
T^{(0)}_{12}T^{(0)}_{21}.\nonumber}
\end{eqnarray}
Substituting (\ref{f11})-(\ref{f15}) into $\mbox{det}_q T(x)$ after
calculations we obtain
$C_{-2}=q^2C_{+2},\ C_0=\alpha^{-1}\alpha_2(1+q)C_{+2},\ C_{\pm 4}=0$
and $C_{+2}=q^{-1}\alpha_2(\lambda^{-1}_3A^{(0)}_+E_{+2}-q^{-1}E^{(0)})K$
that lead to (\ref{g5}) and (\ref{g6}).

It is straightforward to obtain
\begin{eqnarray}
\mbox{tr}T(x) &=&
(A^{(0)}_+ +e^{-i\eta P}A^{(1)}_+ +e^{i\eta P}A^{(2)}_+ +\lambda^{-1}_3
\alpha_2E_{+2}K)x\nonumber\\
&&+(A^{(0)}_- +e^{-i\eta P}A^{(2)}_- +e^{i\eta P}A^{(1)}_- +\lambda^{-1}_3
\alpha_2E_{-2}K^{-1})x^{-1}\nonumber\\
&&+ \lambda_3(e^{-i\eta P}x^3-e^{i\eta P}x^{-3})\label{g9}
\end{eqnarray}
where $A^{(0)}_{\pm}, \ E_{\pm 2}$ are given by (\ref{g1}), whereas
$A^{(1)}_{\pm}$ and  $A^{(2)}_{\pm}$ are given by (\ref{f14}).
In terms of
\begin{equation}
T(x)=\sum^{3}_{n=-3}x^nT^{(n)}=T(u)\sum_{n,m}\sin^nu\cos^muT^{(n,m)}
\label{p6.36} \end{equation}
where $x=e^{iu}$ one finds
\begin{equation}
\mbox{det}_q T(u)=\{2q\cos(2u-\xi\eta)+\alpha^{-1}
\alpha_2(1+q)\}C_{+2} \label{h4}
\end{equation}
and
\begin{eqnarray}
\mbox{tr}T(u) &=& T^{(3,0)}_{11}\sin^3u +T^{(2,1)}_{11}\sin^2u\cos u\nonumber\\
&&+(T^{(1,0)}_{11}+T^{(1,0)}_{22})\sin u+
(T^{(0,1)}_{11}+T^{(0,1)}_{22})\cos u\label{h5}
\end{eqnarray}
Following the Inverse Scattering Methods $T^{(3,0)}_{11}$ and
$T^{(2,1)}_{11}$ correspond to the momentum conservation. The
Hamiltonian is defined by
\begin{equation}
H=f_1\{T^{(1,0)}_{11}+T^{(1,0)}_{22}\} \label{h6}
\end{equation}
and another constant of motion is
\begin{equation}
G=f_2\{T^{(0,1)}_{11}+T^{(0,1)}_{22}\} \label{h7}
\end{equation}
where $f_1$ and $f_2$ are arbitrary constants.

The Hamiltonian of the system given by eqs.~(\ref{h6}) and (\ref{h7}),
take the form
\begin{eqnarray}
H &=& f_1\left( 2i\lambda_3\{\alpha \alpha^{-1}_2(q-1)wg^{-1}\cos[\eta(P+\xi
{\hat J}_3)]{\hat J}_+{\hat J}_-\right.\nonumber\\
&&+2\sin(\xi\eta{\hat J}_3)(\sin[\eta(P+\xi {\hat J}_3]-\alpha^{-1}\alpha_2
q^{\frac 12}\sin[\eta (P+\xi {\hat J}_3+\frac 12)])\nonumber\\
&&+\left.(\alpha\alpha^{-1}_2(q+1)+2)\cos (\xi P)\}+D_+ \right) \label{h8},\\
G &=& f_2\left(-2i\lambda_3\left\{\alpha \alpha^{-1}_2(q-1)wg^{-1}{\hat J}_+
{\hat J}_-+2(\alpha\alpha^{-1}_2q^{\frac 12}\cos[\xi\eta ({\hat J}_3+\frac
12)])
\right. \right.\nonumber\\
&&+\left.\left.\cos (\xi\eta {\hat J}_3)\right\}\sin[\eta(P+\xi
{\hat J}_3]+D_- \right) \label{h9}
\end{eqnarray}
where $D_{\pm}$ are given by
\begin{eqnarray}
D_{\pm} &=& \epsilon_{\pm}\left\{
\lambda^{(1)}_+({\hat J}_-{\hat x}_3+(1-q^{-1})^{-1}q^{-\delta_+}\tau_-
K^{1+\delta_+}(1\mp q^{-1}\alpha^{-1}\alpha_2){\hat x}_-)
\nonumber\right.\\
&&+\lambda^{-1}_3\alpha_2\beta^{(1)}_+(\pm{\hat J}_{\pm}{\hat x}_3
K^{-1}+(1-q^{-1})^{-1}q^{-2+\delta_+}\tau_+\nonumber\\
&&\ \ K^{\delta_+-1}(\alpha^{-1}\alpha_2q^{-1}\mp 1){\hat x}_+)
\left.\right\}\label{h10}
\end{eqnarray}
for $\delta_++\delta_-=1$, and
\begin{eqnarray}
D_{\pm} &=& \epsilon_{\pm}\left\{
\lambda^{-1}_3\alpha_2\beta^{(1)}_+({\hat
J}_{\pm}{\hat x}_3K+(1-q)^{-1}q^{2+\delta_+}\tau_+K^{\delta_++1}(\alpha
\alpha^{-1}_2q\mp 1){\hat x}_+)
\right.\nonumber\\
&&+\lambda^{(1)}_-(\mp{\hat J}_+{\hat x}_3+(1-q)^{-1}q^{-\delta_+}
\tau_-K^{1+\delta_+}(\alpha\alpha^{-1}_2q\mp 1){\hat x}_-)
\left.\right\}\label{h11}
\end{eqnarray}
for $\delta_++\delta_-=-1$, with $\epsilon_+=i,\ \epsilon_-=1$.

It is noted that there are two possibilities:
\begin{equation}
\alpha=\pm q^{-\frac 12}\alpha_2 \label{h12}
\end{equation}
in (\ref{h10}) and (\ref{h11}).

We would like to remark that for the given standard 6-vertex $R$-matrix
eq.~(\ref{p1.21}) we find a set of solution for truncated RTT relation. The
solution can be realized through the algebra (\ref{f18}) and corresponding
conserved quantities (\ref{h8}) and (\ref{h9}). Eq.~(\ref{f18}) naturally
yields the non-commutative geometry~\cite{r12}. Since the considered system is
axially
symmetric so that  the coordinates on $(x_1,\ x_2)$ plane are commute with
each other, whereas the third coordinates $x_3$ does not commute with them.

It can be shown that the rational limit of the trigonometric conserved
quantities
$H$ and $G$ given by eqs.~(\ref{h8}) and (\ref{h9}), respectively, are
exactly the same as the correspondence in the G-C gyrostat model shown by
eqs.~(\ref{e29}) and (\ref{e30}).

All the calculations in this section are tedious and can be found in
Ref.~\cite{xue}.\\

M.-L. Ge and K. Xue would like to thank Prof. Y.M. Cho for his kind invitation,
hospitility and the suggestion for the topic of this papers. The authers
are indebted to Profs. M. Wadati, Y.S. Wu and B.L. Hao for their
stimulating discussions. This work is in part supported by NSF of China.

\end{document}